%% file: main-final.tex
\documentclass[10pt, conference]{IEEEtran}
\IEEEoverridecommandlockouts

\input{preamble}
\input{auto_names}

\newcommand{\measOne}{\roundLabel[black]{Meas 1}}
\newcommand{\measTwo}{\roundLabel[black]{Meas 2}}
\newcommand{\finding}[1]{\emph{#1}}

\usepackage[normalem]{ulem}

\def\BibTeX{{\rm B\kern-.05em{\sc i\kern-.025em b}\kern-.08em
    T\kern-.1667em\lower.7ex\hbox{E}\kern-.125emX}}

\defineauthors[false]{jarne,gilles,liesbet,lieven,benjamin,thomas,todo,geof, reviewcomment, update}

\renewcommand{\update}[1]{#1}

\begin{document}\sloppy

\title{Single versus Multi-Tone Wireless Power Transfer with Physically Large Arrays\thanks{The work is supported by the REINDEER project under grant agreement No.~101013425.}}

\author{
    \IEEEauthorblockN{Jarne Van Mulders\IEEEauthorrefmark{1}, Benjamin J.\,B. Deutschmann\IEEEauthorrefmark{2}, Geoffrey Ottoy\IEEEauthorrefmark{1}, Lieven De Strycker\IEEEauthorrefmark{1},\\ Liesbet Van der Perre\IEEEauthorrefmark{1}, Thomas Wilding\IEEEauthorrefmark{2} and Gilles Callebaut\IEEEauthorrefmark{1}}
    \IEEEauthorblockA{\IEEEauthorrefmark{1}\textit{KU Leuven,} Belgium, 
    \IEEEauthorrefmark{2}\textit{Graz University of Technology,} Austria}
}

\maketitle


\begin{abstract}
Distributed beamforming is a key enabler to provide power wirelessly to a massive amount of \glspl{end}. However, without prior information and fully depleted \glspl{end}, initially powering these devices efficiently is an open question. This work investigates and assesses the feasibility of harvesting sufficient energy to transmit a backscatter pilot signal from the \gls{end}, which can be then used for coherent downlink transmission. We experimentally evaluated adaptive single-tone and multi-tone signals during initial charging. The results indicate that the response time for \glspl{end} with unknown locations can extend to several tens of seconds. Notably, the adaptive single-tone excitation shows, among others, better performance at lower transmit power levels, providing a faster response. These findings underscore the potential of adaptive single-tone signals in optimizing power delivery to \gls{end} in future networks.

\end{abstract}


\glsresetall

\section{Introduction}
\label{section:intro}

\Gls{rfpt} has seen increasing interest from both academia and industry. Applications such as logistic tags and \glspl{esl} can be built without batteries using advanced \gls{wpt} techniques. 
Future 6G networks have the potential to support such applications through distributed \gls{wpt}. 
\Glspl{end} have only limited energy storage available to buffer the required energy in order to provide a stable energy source to the device. Hence, these devices need sufficient power during the initial access phase, i.e., when the energy storage is fully depleted, to charge this buffer.

\textbf{Multi-Antenna \gls{wpt}.}\label{sec:intro-multi-antenna}
Multi-antenna systems offer far-field beamforming and near-field beamfocusing to enhance \gls{wpt} efficiency at targeted devices and reduce unintended power reception. Beamforming typically relies on channel reciprocity, where the infrastructure learns the channel from an uplink pilot signal sent by the device. This information is then used to adjust the antennas' phases and amplitudes to create constructive interference at the target location. However, \glspl{end} need to be powered before they can transmit the necessary pilot signal.

\begin{figure}
    \centering
    \includegraphics[width=0.95\linewidth]{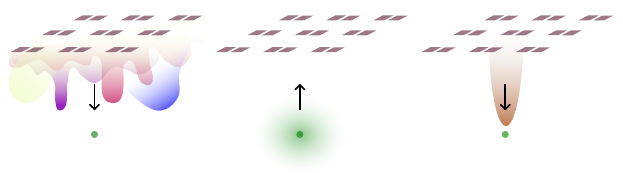}
    \caption{Considered three-phase operation: 1) non-coherent downlink \gls{wpt} (left), 2) 10-byte uplink pilot transmission (middle) and 3) reciprocity-based coherent downlink \gls{wpt} (right). This work focuses on the two first phases.}\label{fig:operation}
\end{figure}

Various methods have been proposed to power devices during initial access, i.e., when the energy storage is fully depleted:

\begin{enumerate}
    \item Uniform Power Distribution: Phases of all antennas are randomly altered over time to achieve a quasi-uniform power distribution, as introduced in~\cite{Lopez21CSIfree-vs-CSIbased,9814679}, which smooths the beamfocusing effect. This yields over time a superposition of the beam patterns as illustrated in phase 1 of \cref{fig:operation}.
    \item Search Algorithms: Algorithms adjust beam weights to maximize received power. For instance, a grid search strategy along the x- and y-axes of the planar array is used in~\cite{10014336}. To reduce search time,~\cite{hajimiriDynamicFocusingLarge2021} employs orthogonal bases to change phases of multiple antenna elements simultaneously, sweeping the phases of antenna groups while recording received power. This technique does not require array calibration or prior knowledge of the environment.
    \item Position-Based \gls{wpt}: Proposed in~\cite{9814679, 10283480}, this method predicts geometry-based \gls{csi} using prior position and environment information. In~\cite{10283480}, geometry-based planar and spherical wavefront beamformers are compared to reciprocity-based beamformers using data from an XL-MIMO testbed.
    \item (Prior) Channel-Based \gls{wpt}: This approach optimizes \gls{wpt} efficiency using previously obtained \gls{csi} or partial \gls{csi}. For instance, \cite{SPAWC2} explores non-coherent and coherent beamforming strategies for initial access using partial \gls{csi}.
\end{enumerate}

\textbf{Single-Tone and Multi-Tone Excitation.}\label{section:sota}
Multi-tone techniques are being proposed in literature as an efficient waveform for \gls{rfpt}. For instance, in~\cite{7932461} the \gls{csi} is exploited to allocate less power to the frequency components (tones) that have weak channel gains. In~\cite{trotter2009power}, the read range of a RFID tag was increased by employing a multi-tone waveform.



However, the aforementioned studies heavily rely on models, not taking into account matching circuit mismatch, different harvester architectures and other hardware impairments, all affecting the performance of the energy harvesting capabilities of an \gls{end}. \update{Moreover, previously mentioned studies rely on partial \gls{csi} knowledge and an optimised phase relation between all multi-tone components, which is not possible in our case (step 1 in \cref{fig:operation}).} More complex AC-to-\acrshort{dc} converters and built-in charge pumps could potentially benefit from single-tone \gls{wpt}. This belief is also challenged in~\cite{shariati2018multitone}, where the impact of multi-tone excitation on the output \gls{dc} power of a voltage-doubler FM-band RF energy harvester is examined. The findings indicate that multi-tone excitation not always enhances the \gls{dc} output power.





\textbf{Contributions and Findings.} 

In this study, we experimentally determine the most effective \update{strategy} for initially charging a real \gls{end} without prior knowledge of its location and with a depleted energy supply. The considered operation is depicted in \cref{fig:operation}. A custom-developed \gls{ep} utilizes a state-of-the-art energy harvesting test IC from NXP. We compare the performance of single-tone and multi-tone transmissions using an \num{84}-antenna system. Our experiments demonstrate that adaptive single-tone transmission outperforms multi-tone waveforms across all key metrics: harvester efficiency (RF-to-\acrshort{dc} conversion), probability of achieving the required \gls{mcu} voltage, and uplink pilot response time. This indicates that single-tone transmission enables \glspl{end} to receive more energy and wake up faster compared to multi-tone signals for initial access. The design files, along with details about the experiment and the collected data, are available in the \faicon{github}~repository~\cite{github}.



 
\section{Large Array, Energy-Neutral Device and Energy Profiler}

\begin{figure}
      \centering
    \includegraphics[width=0.9\linewidth]{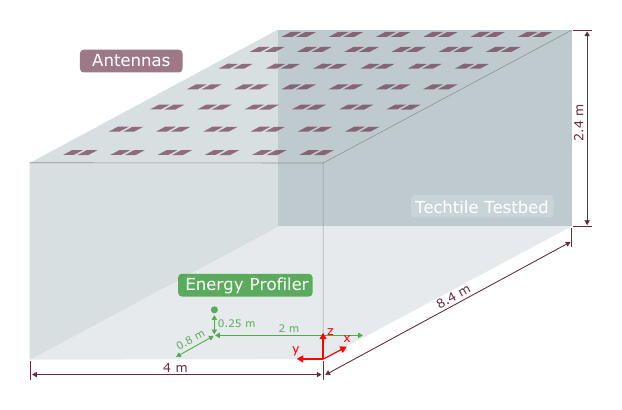}
    \caption{Illustration of the real-life testbed~\cite{callebaut2022techtile} and energy profiler's location.}\label{fig:testbed}
\end{figure}

Performing an empirical comparison between several transmit signals involves numerous hardware and software requirements. This section outlines the developments necessary for conducting practical measurements. Some components are not commercially available and were specifically developed for this work. A flexible and configurable infrastructure is crucial, incorporating large arrays as introduced in \cref{section:testbed}. Additionally, an energy-neutral device implementation is required to determine the necessary energy and energy buffer size, as detailed in \cref{section:energy}. Finally, an \gls{ep} is essential for performing RF-to-\acrshort{dc} conversion and measuring the received \gls{dc} energy, as explained in \cref{section:ep}.

\subsection{Large Array}
\label{section:testbed}

The measurements described in \cref{section:measurements} are performed with the Techtile testbed~\cite{callebaut2022techtile}. Only parts of the infrastructure are used in this work, specifically the antennas located in the ceiling. The setup involves \num{84} antennas paired with \num{42} \gls{usrp} B210 devices. Control of the \glspl{usrp} is managed by \num{42} \glspl{rpi} units, which are connected to the local network of the Techtile testbed. The testbed is time and frequency synchronised, given that multiple \glspl{usrp} could generate exactly the same frequency from a shared \SI{10}{\mega\hertz} clock source. As a result, the relative phase between two frequency synchronous \glspl{usrp} is constant over time. The dimensions of the testbed are illustrated in \cref{fig:testbed}, and a more detailed discussion of the testbed can be found in \cite{callebaut2022techtile}.

\subsection{Energy-Neutral Device and Energy Buffer Requirements}\label{section:energy}

\begin{figure}
     \centering
     \begin{subfigure}[t]{0.9\linewidth}
             \centering
        \includegraphics[width=0.7\linewidth]{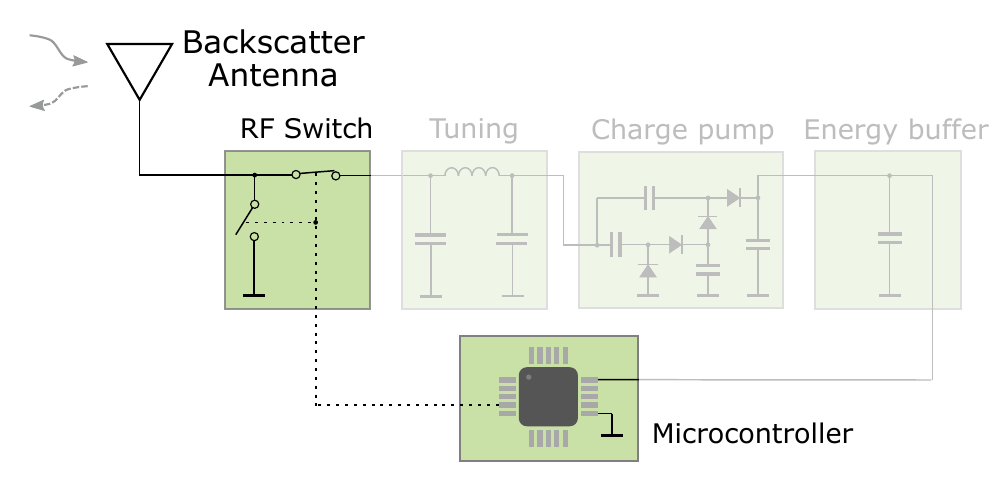}
        \caption{Considered \acrlong{end} design.}
        \label{fig:end_arch}
     \end{subfigure}
     
     \begin{subfigure}[t]{0.79\linewidth}
    \centering
    \includegraphics[width=0.7\linewidth]{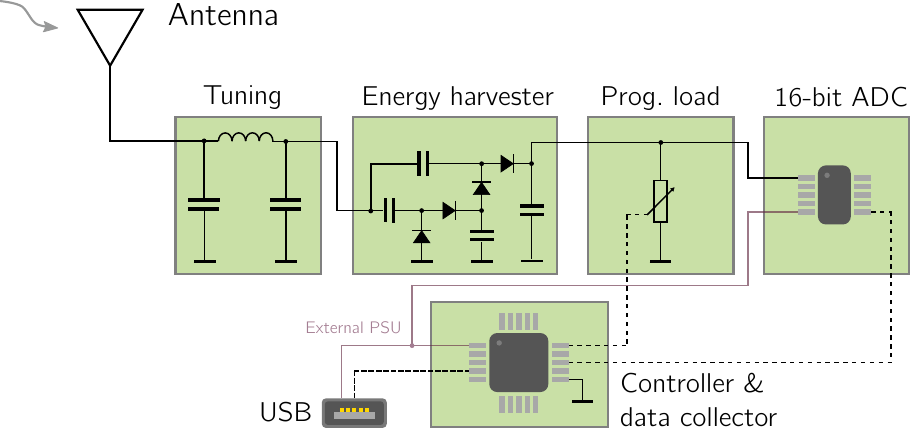}
    \caption{Constructed \acrlong{ep} design.}
    \label{fig:ep-hardware}
    \end{subfigure}
        \caption{Hardware architectures. The transparent blocks of the \gls{end} architecture were not considered during the analysis, as the primary goal was to measure the power consumption of the \gls{mcu} and the RF switch.}
        \label{fig:three graphs}
\end{figure}

For coherent downlink wireless power beamforming, the infrastructure, here in form of a large array, requires \gls{csi}. In order to obtain this, the \gls{end} can use backscattering. The objective of this section is to determine the energy needed at the \gls{end} side to backscatter a pilot signal and, additionally, to ascertain the size of the buffer capacitor.

We consider an embedded device with only backscatter capabilities, as detailed in \cref{fig:end_arch}. A low-power \gls{mcu} uses one of its internal timers to control an external RF switch. The RF switch is located just behind the antenna, allowing it to change the impedance between the harvester impedance and a short-circuit termination.

The timer divides the \SI{1.024}{\mega\hertz} clock by two, to provide a \SI{512}{\kilo\hertz} \gls{lo} offset to the RF switch and therefore prevents interference with the carrier wave. A second timer modulates information at a baud rate of \SI{1}{\kilo\bit\per\second} by controlling the first timer. We assume a pilot message of \num{10} bytes\footnote{\update{Although the pilot length determines the number of \glspl{end} that can be potentially multiplexed, here we do not target simultaneous uplink pilot transmission. Actually, the hypothesis is that using adaptive single carrier transmission, \glspl{end} will awake at different time instances, mitigating pilot contamination. Hence, the pilot length only has an effect on the channel estimate quality, which does not affect the findings and conclusions of this work.} 
}. Consequently, it takes \SI{80}{\milli\second} to backscatter \num{10} bytes of data. The \gls{mcu}, powered at \SI{1.8}{\volt}, consumes \SI{380}{\micro\watt} during operation. Thus, the energy demand of the \gls{mcu} $E_{\text{MCU}}$ amounts \SI{30.4}{\micro\joule}. The data in the message being backscattered is not of interest in this work.

The \gls{mcu} starts executing instructions once the voltage exceeds the threshold value $V_{\text{MCU}_{th}}$ of \SI{1.75}{\volt} and stops executing instructions when the supply voltage drops below the brown-out threshold $V_\text{BOD}$ value of \SI{1.55}{\volt}. Subsequently, the buffer capacitor $C_B$ can be calculated using 

\begin{equation}
    C_\text{B} = \frac{2 E_{\text{MCU}}}{V_{\text{MCU}_{th}}^2 - V_\text{BOD}^2} \,.
    \label{eq:buffercapacitor}
\end{equation}

Inserting the values from above results in a minimum buffer capacitance of \SI{92}{\micro\farad} is required, which means a capacitance of \SI{100}{\micro\farad} should be sufficient to send the pilot message for the considered \gls{end} architecture (\cref{fig:end_arch}).

\subsection{Energy Profiler}
\label{section:ep}

To achieve accurate \gls{dc} power measurements that account for nonlinearities and threshold voltage levels, we have developed and implemented an \acrfull{ep}. The \gls{ep} incorporates several features to validate and optimize the performance of \gls{rfpt} systems. It includes a tuning network that enhances impedance matching with antennas having a characteristic impedance of \SI{50}{\ohm}. The system features an energy harvester that converts RF signals into \gls{dc} energy~\cite{divakaran2019rf}, capable of providing a maximum output voltage of \SI{2}{\volt}, without damaging the internal charge pump circuits. This voltage level should be sufficient to reliably power \glspl{mcu}, as discussed earlier in~\cref{section:energy}. Additionally, the profiler integrates precise power measurement capabilities using a programmable potentiometer and a high-resolution \num{16}-bit \gls{adc}. This setup allows for accurate monitoring of power levels, crucial for evaluating energy harvesting efficiency. Operating at a sample rate and control loop speed of \SI{1}{\kilo\hertz}, the \gls{ep} ensures rapid response times and precise control in dynamic environments. Furthermore, the \gls{ep} offers an adjustable target voltage feature, enabling users to simulate real-life conditions accurately. This flexibility is essential, as \glspl{mcu} typically require a minimum input voltage of approximately \SI{1.8}{\volt} for stable operation. By adjusting the target voltage, the \gls{ep} can simulate varying operating conditions to validate performance under different scenarios effectively. 



\section{Measurement Campaign}
\label{section:measurements}

Our objective is to achieve a rapid response time (phase 2 in \cref{fig:operation}) from the \gls{end}, ensuring that it transmits its pilot message as promptly as possible. Additionally, we seek to accomplish this with minimal overall transmit power from the infrastructure. Three key questions need to be addressed to optimize this process: What is the optimal signal type to apply? How should the \glspl{usrp} be configured? What is the required transmit power per antenna? These questions will be systematically explored through measurements in this section and the subsequent one.

\subsection{Measurement Setup}

\begin{figure}[ht]
    \centering
    \includegraphics[width=0.6\linewidth]{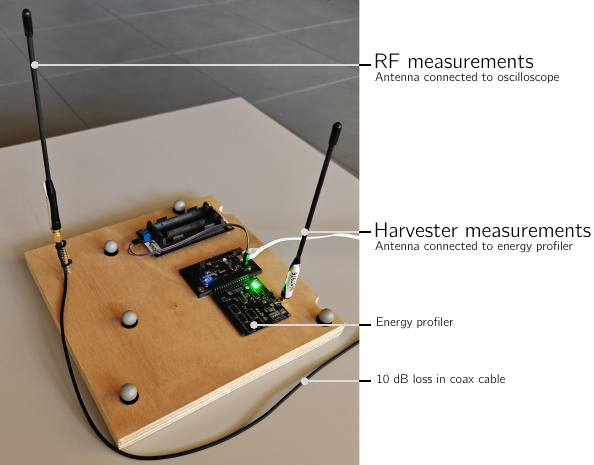}
    \caption{Picture of the receive side of the measurement setup, showing two half-wavelength antennas and the energy profiler.}\label{fig:ep}
\end{figure}

We consider the infrastructure depicted in \cref{fig:testbed}, where the ceiling antennas are utilized for the measurements. In the measurement setup, the receiver has two antennas to compare the harvested \gls{dc} power with the RF channel power, as shown in \cref{fig:ep}. The \gls{ep} connected to the right antenna and explained in \cref{section:ep}, forwards the \gls{dc} power and \gls{dc} voltage values to a central computer at a sample rate of \SI{250}{\hertz}. The target voltage is set to \SI{1.8}{\volt} with an accuracy of \SI{50}{\milli\volt}, meaning that the internal regulator does not adjust the programmable load, if the harvester voltage is within the interval $[1.75, 1.85]$~\si{\volt}. Simultaneously, the RF channel power is measured with the left antenna and an MSO64B oscilloscope (\SI{6}{\giga\hertz} bandwidth) at a sample rate of approximately \SI{20}{\hertz}. 

Furthermore, the measurement instruments, \glspl{rpi}, and \glspl{usrp} are controlled by a central computing system. This system reconfigures the infrastructure for each measurement session and ensures storage of the obtained data.

\subsection{Conducted Measurements}
\label{section:condmeas}

The \glspl{usrp} of Techtile can be configured in various ways. In this paper, we focus on two relevant configurations of the infrastructure:

\measOne{} All transmitters are frequency synchronized and configured to transmit at the same frequency, i.e., \SI{920}{\mega\hertz}. To avoid the possibility of an \gls{end} being located in an area of destructive interference, the phases are adjusted periodically, i.e., adaptive single-tone transmission. In this work, the phase of each antenna is randomly changed every \num{5} seconds.

\measTwo{} The harvester's bandwidth spans several MHz, allowing it to harvest frequency components adjacent to the center frequency of \SI{920}{\mega\hertz}. Consequently, a second configuration involves generating an equally-spaced multi-tone excitation by assigning individual frequencies to each antenna. Here, a frequency offset of \SI{100}{\hertz} is used. \update{Using \num{84} antennas, this equates to \num{84} carriers, spanning a bandwidth of \SI{8.4}{\kilo\hertz}.}\footnote{\update{In contrast to other work, our setup uses single carrier transmission per antenna resulting in a received signal with frequency fading due to different channels between the distributed transmit antennas and the \gls{end} even though a narrowband signal is transmitted.}} 

Since the \gls{end}'s location is often unknown in real-world scenarios, all antennas are configured with the same gain setting for each \gls{usrp}. If the location were known, an optimization algorithm could be employed to efficiently transfer the initial energy to the \gls{end}, as proposed in~\cite{SPAWC2} (\cref{sec:intro-multi-antenna}).

The transmit gain of the \num{84} antennas is swept from \SI{75}{\dB} to \SI{85}{\dB} in \SI{1}{\dB} increments, corresponding to a real output power range of \num{9.1} to \SI{17.4}{\dBm} per antenna. This varying gain is expected to demonstrate the nonlinearities of the harvester, allowing to determine the required gain (or equivalently transmit power) necessary to provide sufficient initial energy to the \gls{end}. In total, \num{11} gain configurations are tested per measurement, each with a duration of \num{30} minutes. 

All measurements are conducted at a static location (\cref{fig:testbed}). It is important to note that the locations of the two antennas are not identical, thus the antennas will experience different small-scale fading. However, the average received RF and \gls{dc} powers are still relevant for comparison purposes, as discussed in \cref{section:processing}. The location of the receiver in our measurement setup remained unchanged in this study (\cref{fig:testbed}). 

\section{Data Processing and Discussion}
\label{section:processing}

In this section, we demonstrate that for the same amount of radiated power, single-tone excitation yields better harvester efficiency and increases the likelihood of receiving a response. A detailed discussion is provided in this section.




\subsection{Average RF/\acrshort{dc} Energy and Harvester Efficiency}

\begin{figure*}[ht]
    \centering
    \input{tikz/harv-efficiency}
    \caption{Average power levels and corresponding efficiency levels of the harvester. 
    The plot illustrates the relationship between average RF input- and \gls{dc} output-power across the energy harvester. The adaptive single-tone configuration achieves, on average, higher efficiency levels.}   
    \label{fig:harveffrealmeas}
\end{figure*}
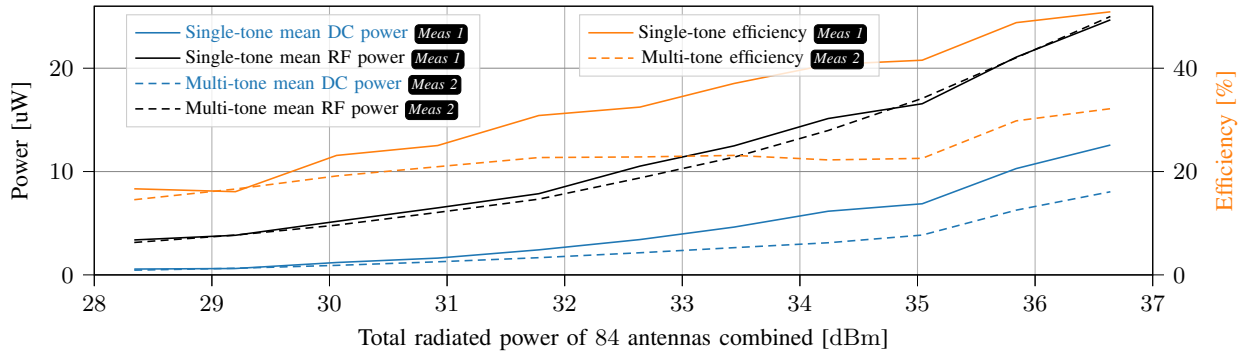

Due to the distance between the scope and \gls{ep} antenna, the received power is different at both locations due to small-scale fading effects, as previously mentioned in \cref{section:condmeas}. For each individual measurement, such as those shown in \cref{fig:buf_single_tone_vs_time,fig:buf_multi_tone_vs_time}, the average RF and \gls{dc} power is determined over the entire measurement duration. The relationship between these two results indicates the harvester efficiency. The average RF and \gls{dc} power levels, as well as efficiency metrics for both single-tone and multi-tone excitation, are plotted in \cref{fig:harveffrealmeas}.

The results from our experiments indicate that single-tone excitation yields higher efficiency values compared to multi-tone excitation. This improvement is attributable to the higher measured average \gls{dc} power (per \gls{usrp} gain) observed for single-tone signals. As expected, the average RF power (represented by the black lines) remains approximately constant across both excitation modes. This consistency arises because the differences between the transmitted signals---namely, the frequency offsets and periodic phase
adjustments---do not cause significant variations in the average RF channel power.


In these measurements, the total transmitted power is defined by the number of antennas and the transmit power. The overall efficiency is presented in \cref{tab:toteff_targvolt}.


\begin{table}[ht]
    \centering
    \caption{Overall efficiency levels and feasibility of the target voltage for various transmit power levels, with $V_{th}$ equals $V_{\text{MCU}_{th}}$.}
    \resizebox{0.48\textwidth}{!}{
    \begin{tabular}{ccc||cc||cc}
        \toprule
        \multicolumn{2}{c}{USRP} & Total & \multicolumn{2}{c||}{Single-tone signal}  &  \multicolumn{2}{c}{Multi-tone signal} \\
        
        gain & power & power &  Efficiency & $V_{\text{B}} > V_{th}$ & Efficiency & $V_{\text{B}} > V_{th}$ \\
       
        [\si{\dB}] & [\si{\dBm}] & [\si{\dBm}] & [\si{\ppm}] &  [\si{\percent}] & [\si{\ppm}] &  [\si{\percent}] \\ 
        \midrule
        75 & 9.1    & 28.34    & 0.8 & 2.5    & 0.7 & 0.0    \\
        76 & 9.96   & 29.2    & 0.8 & 2.77   & 0.8 & 0.0    \\
        77 & 10.82  & 30.06  & 1.2 & 8.01   & 0.9 & 0.0    \\
        78 & 11.68  & 30.92  & 1.3 & 11.54  & 1.0 & 0.59    \\
        79 & 12.54  & 31.78  & 1.6 & 18.87  & 1.1 & 1.93   \\
        80 & 13.4   & 32.64   & 1.9 & 27.59  & 1.2 & 14.98  \\
        81 & 14.2   & 33.44   & 2.1 & 37.52  & 1.2 & 31.39  \\
        82 & 15.0   & 34.24   & 2.3 & 43.01  & 1.2 & 50.66  \\
        83 & 15.8   & 35.04   & 2.2 & 44.8   & 1.2 & 57.34  \\
        84 & 16.6   & 35.84   & 2.7 & 53.69  & 1.6 & 57.95  \\
        85 & 17.4   & 36.64   & 2.7 & 62.08  & 1.7 & 53.91  \\
        \bottomrule
    \end{tabular}}
    \label{tab:toteff_targvolt}
\end{table}


\subsection{Feasibility of the Target Voltage}
One of the key questions is whether the harvested \gls{dc} power provides sufficient voltage for the \gls{mcu} to wake-up and stay active long enough to transmit an uplink pilot. Therefore, the initial study evaluates whether the harvested voltage level exceeds the \SI{1.75}{\volt} threshold ($V_{\text{MCU}_{th}}$). This voltage level is measured by the \gls{ep}.
\Cref{tab:toteff_targvolt} indicates that at lower \gls{usrp} gain levels, the harvested voltage in multi-tone measurements does not consistently exceed $V_{\text{MCU}_{th}}$. In contrast, single-tone signals exhibit a higher probability of surpassing this threshold, particularly at lower transmit power levels.

\subsection{Estimation of the END Response Time}

One of the main contributions of this manuscript is to provide insights into the response time of the \gls{end}. Based on this information, it is feasible to estimate how long the infrastructure needs to remain actively listening for backscattered signals. \Cref{section:energy} defined an \gls{end} buffer of $\SI{100}{\micro\farad}$ that should be charged to $V_{\text{MCU}_{th}}$ to transmit the pilot message. To proceed with the response time calculations, the voltage profile across the buffer must be known. \cref{section:responstime} explains how the results from \cref{section:measurements} can be processed to estimate the voltage profile across the capacitor. \Cref{section:analysis} analyzes whether the buffer voltage effectively reaches the desired voltage levels.

\subsubsection{Estimated Time to Charge an \gls{end} Buffer}
\label{section:responstime}

The buffer voltage $V_\text{B}(t)$ is calculated according to 

\begin{equation}
V_\text{B} (t) = \sum_{n=1}^{t/\Delta t} V_{\text{B}_n} \quad \text{for} \quad t/\Delta t \leq N
\label{eq:sumbuffervoltage}
\end{equation}


with $V_{\text{B}_n}$ the voltage after processing $n$ samples and $N$ the number of data samples available in the \gls{ep} dataset. The time interval $\Delta t$ depends on the \gls{ep} sampling rate. The instantaneous change in energy $\Delta E_{\text{B}_n}=P_{\text{B}_n} \Delta t$ is computed according to

\begin{equation}
    \Delta E_{\text{B}_n} = \frac{1}{2} C_B \left(V_{\text{B}_n}^2 - V_{\text{B}_{n-1}}^2\right)
    \label{eq:instenergy}\,.
\end{equation}

\begin{figure}
    \centering
    \begin{subfigure}[b]{\columnwidth}
        \centering
        \resizebox{0.9\columnwidth}{!}{\input{tikz/one_tone_phase_duration_5_m1_80_energy_buffer}}
        \caption{\measOne{} Single-tone excitation.}
        \label{fig:buf_single_tone_vs_time}
    \end{subfigure}
    \hfill
    \vspace{0.5mm}
    \begin{subfigure}[b]{\columnwidth}
        \centering
        \resizebox{0.9\columnwidth}{!}{\input{tikz/multi_tone_m1_80_energy_buffer}}
        \caption{\measTwo{} Multi-tone excitation.}
        \label{fig:buf_multi_tone_vs_time}
    \end{subfigure}
    \caption{Buffer voltage progression with \acrshort{usrp} gain of \SI{80}{\dB}, corresponding to a total transmit power of \SI{32.64}{\dBm} (\cref{tab:toteff_targvolt}). 
    for \measTwo{}, the buffer never reaches the target voltage. }
    \label{fig:buf_signals_vs_time}
\end{figure}
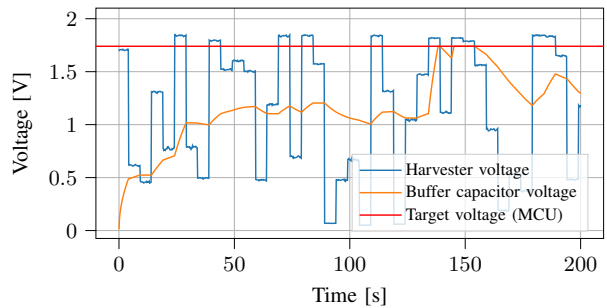
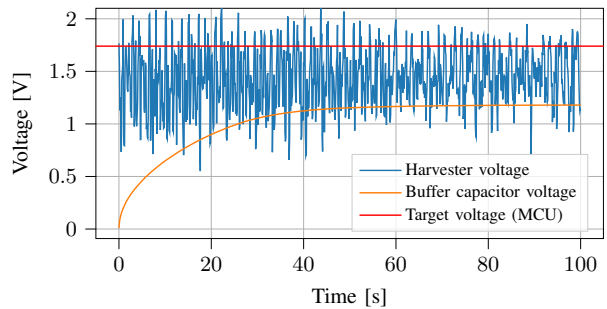

The total power supplied to or consumed by the buffer capacitor $P_{\text{B}_n}$ is defined by the harvested power $P_{\text{EH}_n}$ minus the power drawn by the load $P_{\text{MCU}_n}$. The load in this application includes only the \gls{mcu}, and is related to the supply voltage $V_{\text{MCU}}$. It is essential to recognize that the \gls{mcu} still consumes several microwatts of power at voltages below the threshold voltage $V_{\text{MCU}_{th}}$. If the voltage of \SI{1.75}{\volt} is exceeded, $P_{\text{MCU}}$ increases and the \gls{mcu} starts executing instructions.
Based on the power $P_{\text{B}_n} = P_{\text{EH}_n} \!-\! P_{\text{MCU}_n}$, the buffer capacitance $C_{\text{B}}$, and the voltage across the buffer capacitor at the previous time step $V_{\text{B}_{n-1}}$, the voltage $V_{\text{B}_n}$ can be determined via 


\begin{equation}
    V_{\text{B}_n} = 
    \begin{cases}
    0 , & V_{\text{B}_n} < 0\\
    V_{\text{B}_{n-1}} , & V_{\text{EH}_n} > V_{\text{B}_n}, P_{\text{B}_n} > 0 \\
    \sqrt{V_{\text{B}_{n-1}}^2 + \frac{2P_{\text{B}_n}\Delta t}{C_{\text{B}}}} , & \text{else}\\
    \end{cases}
    \label{eq:vbn}
\end{equation}

with $P_{\text{EH}_n}$ depending on the incoming RF power $P_{\text{RF}_{\text{in}}}$ and $P_{\text{MCU}_n}$ depending on $V_{\text{B}_{n-1}}$. The \update{energy harvester output} voltage $V_{\text{EH}}$ changes over time and can drop below the voltage already present across the buffer capacitor, i.e., $V_{\text{EH}_n} < V_{\text{B}_n}$. 
When $V_{\text{EH}_n} > V_{\text{B}_n}$ and $P_{\text{B}_n} > 0$, the harvester is directly powering the \gls{mcu}, keeping the buffer voltage unaltered as the voltage $V_\text{B}$ cannot increase further due to an insufficient $V_{\text{EH}}$, i.e., $V_{\text{B}_n} = V_{\text{B}_{n-1}}$.
Once the buffer voltage $V_{\text{B}_n}$ reaches the \gls{mcu} threshold voltage, sufficient energy is stored in the buffer to backscatter the pilot message. Furthermore, the response time can be estimated by calculating the buffer voltage progression. The calculations were applied to the measured data, available in~\faicon{github}~repository~\cite{github}, resulting in the representations of the buffer voltage progression shown in \cref{fig:buf_single_tone_vs_time,fig:buf_multi_tone_vs_time} and discussed in detail below.\footnote{It may be observed that these calculations are not strictly necessary, as the buffer capacitance can be directly connected to the energy harvester. By consequently monitoring the buffer voltage over time, similar results can be obtained. However, the proposed approach facilitates a more comprehensive post-processing analysis. 
Suppose the pilot message, capacitor value or \gls{mcu} load is changed, re-running the analyses immediately provides insights into the feasibility, charge time, and response time of the \gls{end}.}

\subsubsection{Response Time Results}
\label{section:analysis}

To obtain a realistic estimate of the response time, a Monte-Carlo simulation is run with the measured data. \num{50} random time instances are selected from the dataset and the time to charge the buffer is calculated according to \cref{section:responstime}. Given that the \gls{mcu} consumes power below the threshold, two configurations are considered, i.e., realistic\footnote{The power consumption at different supply voltages of the \gls{mcu} is measured and are available in~\cite{github}.} and ideal \gls{mcu}. In the ideal case, the \gls{mcu} does not consume power below the threshold voltage. This specific case affect the response time calculations and shows that future \glspl{end} incorporating reduced \glspl{mcu} power consumption during periods of inactivity will have a significant impact on the results.


The 50th percentile (P50) and 95th percentile (P95) values for the single-tone (\measOne{}) and multi-tone (\measTwo{}) transmission for both the realistic and ideal \gls{mcu} are summarized in \cref{tab:response_time_single_multi} for different total transmit powers. From this, we can conclude that \finding{for lower transmit powers, and in contrast to multi-tone, single-tone signals exceed the required threshold values, allowing to send a pilot signal}. Furthermore, \cref{tab:response_time_single_multi} also shows that \finding{lowering the power consumption of the \gls{mcu} below $V_{\text{MCU}_{th}}$ allows to reduce the overall transmit power}. 

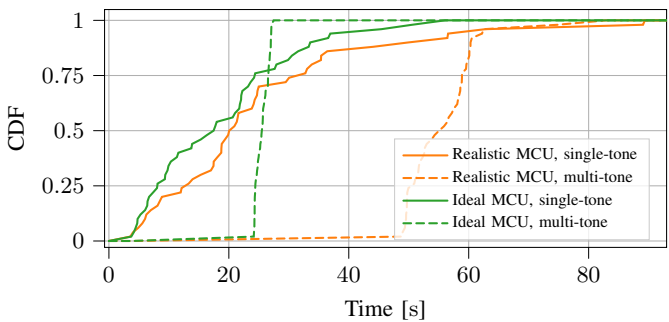
\begin{figure}[ht]
    \centering
    \setlength{\figurewidth}{0.85\linewidth}%
    \setlength{\figureheight}{0.45\linewidth}   
    \resizebox{1\columnwidth}{!}{\input{tikz/cdf}}
        \caption{\Gls{cdf} of the response times for a total transmit power of \SI{35.84}{\dBm}.}
    \label{fig:cdf}
\end{figure}

\Cref{fig:cdf} plots the \gls{cdf} of the response times over the \num{50} Monte-Carlo measurement-based simulations. It shows, that \finding{while the single-tone has a higher spread of response times, it also has a higher probability of experiencing a shorter response time than the multi-tone signal}. This can be understood when inspecting \cref{fig:buf_signals_vs_time}, where multi-tone signals provide a more stable received \gls{dc} power, while the adaptive single-tone exploits constructive interference over a longer time period as can be observed in \cref{fig:buf_single_tone_vs_time}.

\begin{table}[ht]
    \centering
    \caption{Computed 50th (P50) and 95th percentile (P95) of the \gls{end} response time derived from single- and multi-tone measurements. Some configurations result in a response time exceeding the measurement time of \SI{1}{\hour}, indicated by 
 -.}\label{tab:response_time_single_multi}
    \resizebox{0.48\textwidth}{!}{
    \begin{tabular}{c||rr|rr||rr|rr}
        \toprule
        Total & \multicolumn{4}{c||}{Single-tone} & \multicolumn{4}{c}{Multi-tone} \\
        transmit & \multicolumn{2}{c|}{Realistic MCU} & \multicolumn{2}{c||}{Ideal MCU} & \multicolumn{2}{c|}{Realistic MCU} & \multicolumn{2}{c}{Ideal MCU} \\
        power & P50 & P98 & P50 & P98 & P50 & P98 & P50 & P98 \\
        
        [\si{\dBm}] & [\si{\second}] & [\si{\second}] & [\si{\second}] & [\si{\second}] & [\si{\second}] & [\si{\second}] & [\si{\second}] & [\si{\second}] \\
        \midrule
        28.34  &  -   &  -    & 510 & 1159 &  -     & -     & -      &  -     \\
        29.20  &  -   & -     & 418 & 756  &  -     & -     & -      &  -     \\
        30.06  & -    &   -   & 206 & 389  &  -     & -     & -      &  -     \\
        30.92  & 620 & 1537 & 132 & 310  & -     & -     & 820    & 848 \\
        31.78  & 156 & 483  & 84  & 166  & -     & -     & 288    & 328 \\
        32.64  & 148 & 410  & 53  & 127  & -     & -     & 102    & 112 \\
        33.44  & 76  & 438  & 39  & 102  & -     & -     & 69     & 71  \\
        34.24  & 55  & 221  & 28  & 67   & -     & -     & 53     & 54  \\
        35.04  & 31  & 125  & 23  & 75   & -     & -     & 41     & 48  \\
        35.84  & 20  & 89   & 17  & 50   & 55    & 74    & 25     & 27  \\
        36.64  & 17  & 80   & 14  & 38   & 31    & 43    & 20     & 20  \\
        \bottomrule
    \end{tabular}}
\end{table} 




\section{Conclusion}

In this paper, we report and discuss various measurements at a static location within the Techtile testbed. It was found that, for the employed energy harvester, single-tone excitation yields the best results in terms of efficiency levels and \gls{end} response time. A fair comparison was achieved by evaluating the results using the same amount of radiated power. However, it was also observed that the response time amounts to several tens of seconds to backscatter just a \num{10}-byte pilot signal. 
Future work should explore \update{the analytical verification to prove that single-tone signals perform better in this particular case. Furthermore, the measurements will be repeated and multi-carrier transmission per antenna will be considered, as opposed to the current implementation. Also,} the consumption of the inactive \gls{mcu} could be reduced, as we have demonstrated that lowering the threshold voltage for the \gls{mcu} leads to substantial performance improvements. 
There is also a broad spectrum of alternative options to the pure single-tone and multi-tone signals. Future investigations could explore whether different configurations of the infrastructure and, e.g., number of carriers, could yield performance gains. Moreover, the \gls{ep} could be enhanced by integrating the buffer capacitor, allowing for more accurate validation of response time estimations.



The next step to support coherent downlink wireless power beamforming is capturing the backscatter uplink pilot signals and performing reciprocity-based downlink \gls{wpt} beamforming \update{(step 3 in \cref{fig:operation})}.


\printbibliography
\end{document}

%% file: preamble.tex
\usepackage{graphicx} 
\graphicspath{{./figures}}
\usepackage[OT1]{fontenc}
\usepackage{color}
\usepackage[dvipsnames,svgnames, table]{xcolor} 
\usepackage{siunitx}
\DeclareSIUnit{\dB}{dB}	                
\DeclareSIUnit{\dBm}{dBm}	                
\DeclareSIUnit{\dBi}{dBi}                   
\DeclareSIUnit{\dBsm}{dBsm}                 
\DeclareSIUnit{\ppm}{ppm}
\usepackage{booktabs}
\usepackage[backend=biber, style=ieee, citestyle=numeric-comp, maxcitenames=1,mincitenames=1, maxbibnames=1, minbibnames=1,isbn=false, doi=false,citecounter=true]{biblatex}

\usepackage{circuitikz}

\usepackage{pgfplots}
\usepgfplotslibrary{groupplots,dateplot}
\usetikzlibrary{patterns,shapes.arrows}
\usetikzlibrary{spy}
\pgfplotsset{compat=newest}
\usetikzlibrary{external}
\usetikzlibrary{calc}
\usetikzlibrary{shadings}
\usetikzlibrary{shadows.blur}
\usetikzlibrary{decorations.pathreplacing}

\usepackage{soul}

\pgfmathdeclarefunction{viridis}{1}{%
  \pgfmathparse{%
    1000*(
    0.370489*pow(#1, 3)-
    1.11384*pow(#1, 2)+
    0.929777*(#1)
    )%
  }%
}

\usepackage{tabularx}
\usepackage{tablefootnote}

\AtBeginBibliography{\footnotesize}

\usepackage[font=small]{caption}
\usepackage{subcaption}
\usepackage{xspace}

\usepackage{placeins}
\usepackage[xindy, nonumberlist]{glossaries}
\makenoidxglossaries%
\loadglsentries{abbr}%

\usepackage{comment}

\newcolumntype{R}{>{\raggedleft\arraybackslash}X}

\newcolumntype{H}{>{\setbox0=\hbox\bgroup}c<{\egroup}@{}}

\usepackage{graphicx}
\usepackage{textcomp}
\usepackage{xcolor}
\def\BibTeX{{\rm B\kern-.05em{\sc i\kern-.025em b}\kern-.08em
    T\kern-.1667em\lower.7ex\hbox{E}\kern-.125emX}}

\usepackage[capitalize]{cleveref}

 \usepackage{fontawesome}

\pgfplotsset{
    every axis/.append style={
        legend style={
            font=\small,       
            nodes={scale=0.8, transform shape}
        },
         xlabel style={
            font=\small        
        },
        ylabel style={
            font=\small        
        },
    },
    every tick label/.append style={font=\small}
}

\tikzset{
    every node/.append style={font=\footnotesize},
    every label/.append style={font=\footnotesize}
}
\pgfplotsset{
    every axis legend/.append style={
        fill opacity=0.8,
        draw=none,
    },
    every axis/.append style={
        x grid style={darkgray!60},
        xmajorgrids,
        y grid style={darkgray!60},
        ymajorgrids,
    }
}

\newlength{\figurewidth}
\newlength{\figureheight}

\usepackage[left=0.625in,right=0.625in,top=0.75in, bottom=1in]{geometry}

%% file: abbr.tex

\newacronym{2d}{2D}{two-dimensional}
\newacronym{3d}{3D}{three-dimensional}
\newacronym{3gpp}{3GPP}{3rd generation partnership project}
\newacronym{4g}{4G}{fourth-generation}
\newacronym{5g}{5G}{fifth-generation}
\newacronym{6dof}{6DoF}{six degrees of freedom}
\newacronym{6g}{6G}{sixth-generation}
\newacronym{abp}{ABP}{authentication by personalisation}
\newacronym{ac}{AC}{alternating current}
\newacronym{aclr}{ACLR}{adjacent channel leakage ratio}
\newacronym{adc}{ADC}{analog-to-digital converter}
\newacronym{adr}{ADR}{adaptive data rate}
\newacronym{aes}{AES}{Advanced Encryption Standard}
\newacronym{afe}{AFE}{analog front end}
\newacronym{ag}{AG}{array gain}
\newacronym{agps}{A-GPS}{Assisted Global Positioning System} 
\newacronym{ai}{AI}{artificial intelligence}
\newacronym{alk}{ALK}{Alkaline}
\newacronym{am}{AM}{amplitude modulation}
\newacronym{amam}{AM/AM}{amplitude modulation to amplitude modulation}
\newacronym{amc}{AMC}{automatic modulation classification}
\newacronym{amp}{AMP}{approximate message passing}
\newacronym{ampm}{AM/PM}{amplitude modulation to phase modulation}
\newacronym{aoa}{AOA}{angle-of-arrival}
\newacronym{aod}{AOD}{angle-of-departure}
\newacronym{ap}{AP}{access point}
\newacronym{api}{API}{application program interface}
\newacronym{apt}{APT}{acoustic power transfer}
\newacronym{apu}{APU}{access point unit}
\newacronym{ar}{AR}{augmented reality}
\newacronym{arima}{ARIMA}{Auto-Regressive Integrated Moving Average}
\newacronym{arp}{ARP}{Antenna Reference Point}
\newacronym{asic}{ASIC}{application specific integrated circuit}
\newacronym{ask}{ASK}{amplitude-shift keying}
\newacronym{at}{AT}{ATtention}
\newacronym{auv}{AUV}{autonomous underwater vehicle}
\newacronym{awg}{AWG}{arbitrary waveform generator}
\newacronym{awgn}{AWGN}{additive white Gaussian noise}
\newacronym{baw}{BAW}{bulk acoustic wave}
\newacronym{bb}{BB}{base-band}
\newacronym{bc}{BC}{backscatter communication}
\newacronym{bd}{BD}{backscatter device}
\newacronym{be}{BE}{Belgium}
\newacronym{ber}{BER}{bit error rate}
\newacronym{bf}{BF}{beamforming}
\newacronym{bga}{BGA}{ball grid array}
\newacronym{bldc}{BLDC}{brushless DC}
\newacronym{bod}{BOD}{Brown-Out Detection}
\newacronym{bom}{BOM}{bill of materials}
\newacronym{bpsk}{BPSK}{binary phase-shift keying}
\newacronym{bs}{BS}{base station}
\newacronym{bu}{BU}{booster unit}
\newacronym{bw}{BW}{bandwidth}
\newacronym{ca}{CA}{carrier emitter}
\newacronym{cad}{CAD}{channel activity detection}
\newacronym{cars}{CARS}{calibration reference signal}
\newacronym{cbm}{CBM}{condition based maintenance}
\newacronym{cc}{CC}{constant current}
\newacronym{ccdf}{CCDF}{complementary cumulative distribution function}
\newacronym{ccnn}{CCNN}{circular convolutional neural network}
\newacronym{ccs}{CCS}{correlative channel sounder}
\newacronym{cdf}{CDF}{cumulative distribution function}
\newacronym{cdma}{CDMA}{code division-multiple access}
\newacronym{cdrx}{CDRX}{connected mode DRX}
\newacronym{ce}{CE}{coverage enhancement}
\newacronym{ced}{CED}{cumulative energy density}
\newacronym{cf}{CF}{cell-free}
\newacronym{cfmmimo}{CF-mMIMO}{cell-free massive MIMO}
\newacronym{cfo}{CFO}{carrier frequency offset}
\newacronym{cir}{CIR}{channel impulse response}
\newacronym{cla}{CLA}{closed-loop approach}
\newacronym{clk}{CLK}{clock}
\newacronym{cmos}{CMOS}{complementary metal oxide semiconductor}
\newacronym{cost}{COST}{commercial off-the-shelf}
\newacronym{cots}{COTS}{commercial off-the-shelf}
\newacronym{cp}{CP}{cyclic prefix}
\newacronym{cpt}{CPT}{capacitive power transfer}
\newacronym{cpu}{CPU}{central-processing unit}
\newacronym{cpw}{CPW}{coplanar waveguide}
\newacronym{cqi}{CQI}{channel quality indicator}
\newacronym{cr}{CR}{coding rate}
\newacronym{crc}{CRC}{cyclic redundancy check}
\newacronym{crlb}{CRLB}{Cram\'er-Rao lower bound}
\newacronym{crs}{CRS}{cell reference signal}
\newacronym{cs}{CS}{compressed sensing}
\newacronym{cse}{CSE}{channel state estimation}
\newacronym{csi}{CSI}{channel state information}
\newacronym{csp}{CSP}{contact service point}
\newacronym{css}{CSS}{chirp spread spectrum}
\newacronym{cu}{CU}{central unit}
\newacronym{cv}{CV}{constant voltage}
\newacronym{cw}{CW}{continuous wave}
\newacronym{d2d}{D2D}{device-to-device}
\newacronym{dab}{DAB}{Digital Audio Broadcasting}
\newacronym{dac}{DAC}{digital-to-analog converter}
\newacronym{daq}{DAQ}{data acquisition system}
\newacronym{das}{DAS}{distributed antenna systems}
\newacronym{dbpsk}{DBPSK}{Differential Binary Phase Shift Keying}
\newacronym{dc}{DC}{direct current}
\newacronym{dcc}{DCC}{dynamic cooperation clustering}
\newacronym{ddc}{DDC}{digital down conversion}
\newacronym{de}{DE}{drain efficiency}
\newacronym{dft}{DFT}{discrete Fourier transform}
\newacronym{dl}{DL}{downlink}
\newacronym{dlc}{DLC}{Distributed Laser Charging}
\newacronym{dli}{DLI}{direct link interference}
\newacronym{dlt}{DLT}{Distributed Ledger Technology}
\newacronym{dm}{DM}{diffuse multipath}
\newacronym{dma}{DMA}{Direct Memory Access}
\newacronym{dmac}{DMAC}{Direct Memory Access Controller}
\newacronym{dmc}{DMC}{diffuse multipath component}
\newacronym{dmimo}{D-MIMO}{distributed MIMO}
\newacronym{doa}{DOA}{direction-of-arrival}
\newacronym{dp}{DP}{differencial privacy}
\newacronym{dpd}{DPD}{digital pre-distortion}
\newacronym{dpdk}{DPDK}{Data Plane Development Kit}
\newacronym{dram}{DRAM}{dynamic random-access memory}
\newacronym{drcs}{$\Delta$RCS}{differential-radar cross section}
\newacronym{drx}{DRX}{Discontinuous Reception Mode}
\newacronym{dsb}{DSB}{double-sideband}
\newacronym{dsl}{DSL}{digital subscriber line}
\newacronym{dsp}{DSP}{digital signal processing}
\newacronym{dss}{DSS}{Dataset Storage Standard}
\newacronym{duc}{DUC}{digital up-converter}
\newacronym{dvb}{DVB}{Digital Video Broadcasting}
\newacronym{e2e}{E2E}{end-to-end}
\newacronym{easa}{EASA}{European Union Aviation Safety Agency}
\newacronym{ebg}{EBG}{electromagnetic bandgap}
\newacronym{ec}{EC}{European Commission}
\newacronym{ecc}{ECC}{elliptic curve cryptography}
\newacronym{ecdf}{eCDF}{empirical cumulative distribution function}
\newacronym{ecdlp}{ECDLP}{elliptic curve discrete logarithm problem}
\newacronym{ecsp}{ECSP}{edge computing service point}
\newacronym{edlc}{EDLC}{electrostatic double-layer capacitors}
\newacronym{edrx}{eDRX}{Extended Discontinuous Reception Mode}
\newacronym{ee}{EE}{energy efficiency}
\newacronym{egprs}{EGPRS}{Enhanced Data Rates for GSM Evolution}
\newacronym{eh}{EH}{energy harvesting}
\newacronym{eirp}{EIRP}{equivalent isotropically radiated power}
\newacronym{em}{EM}{electromagnetic}
\newacronym{embb}{eMBB}{enhanced Mobile Broadband}
\newacronym{en}{EN}{energy neutral}
\newacronym{end}{END}{energy-neutral device}
\newacronym{enob}{ENOB}{effective number of bits}
\newacronym{eol}{EoL}{end of life}
\newacronym{epu}{EPU}{edge processing unit}
\newacronym{er}{ER}{Energy Receiver}
\newacronym{erp}{ERP}{effective radiation power}
\newacronym[plural=ESCs,firstplural=electronic speed controllers (ESCs)]{esc}{ESC}{electronic speed control}
\newacronym{esd}{ESD}{electrostatic discharge}
\newacronym{esl}{ESL}{electronic shelf label}
\newacronym{esr}{ESR}{equivalent series resistance}
\newacronym{et}{ET}{Energy Transmitter}
\newacronym{etsi}{ETSI}{European Telecommunications Standards Institute}
\newacronym{evd}{EVD}{eigenvalue decomposition}
\newacronym{evm}{EVM}{Error Vector Magnitude}
\newacronym{ewlb}{eWLB}{Embedded Wafer Level Ball Grid Array}
\newacronym{fa}{FA}{federation anchor}
\newacronym{fair}{FAIR}{Findability, Accessibility, Interoperability, and Reuse of digital assets}
\newacronym{fd}{FD}{front-haul distance}
\newacronym{fde}{FDE}{frequency domain equalizer}
\newacronym{fdm}{FDM}{frequency-division multiplexing}
\newacronym{fdma}{FDMA}{frequency division multiple access}
\newacronym{fem}{FEM}{finite element analysis}
\newacronym{fembb}{feMBB}{further enhanced mobile broadband}
\newacronym{fft}{FFT}{fast Fourier transform}
\newacronym{fh}{FH}{fronthaul}
\newacronym{fhss}{FHSS}{frequency hopping spread spectrum}
\newacronym{fifo}{FIFO}{First In, First Out}
\newacronym{fim}{FIM}{Fisher information matrix}
\newacronym{fits}{FITS}{Flexible Image Transport System}
\newacronym{fl}{FL}{federated learning}
\newacronym{fom}{FoM}{figure of merit}
\newacronym{fpga}{FPGA}{field-programmable gate array}
\newacronym{fram}{FRAM}{Ferroelectric Random Access Memory}
\newacronym{fsk}{FSK}{frequency shift keying}
\newacronym{fss}{FSS}{frequency selective surface}
\newacronym{gb}{GB}{grant-based}
\newacronym{gdpr}{GDPR}{general data protection regulation}
\newacronym{gf}{GF}{grant-free}
\newacronym{gmsk}{GMSK}{Gaussian minimum-shift keying}
\newacronym{gnb}{gNB}{Next Generation Node B}
\newacronym{gni}{GNI}{gross national income}
\newacronym{gnn}{GNN}{graph neural network}
\newacronym{gnss}{GNSS}{global navigation satellite system}
\newacronym{gpclk}{GPCLK}{general purpose clock}
\newacronym{gpio}{GPIO}{General-Purpose Input/Output}
\newacronym{gpl}{GPL}{GNU General Public License}
\newacronym{gprs}{GPRS}{General Packet Radio Services}
\newacronym{gps}{GPS}{Global Positioning System}
\newacronym{gpu}{GPU}{graphical processing unit}
\newacronym{grc}{GRC}{GNU Radio Companion}
\newacronym{gscm}{GSCM}{geometry‐based stochastic model}
\newacronym{gsm}{GSM}{Global System for Mobile Communications}  %
\newacronym{gwp}{GWP}{Global Warming Potential}
\newacronym{harq}{HARQ}{hybrid automatic repeat request}
\newacronym{hat}{HAT}{hardware attached on top}
\newacronym{hcs}{HCS}{human-centric services}
\newacronym{hdf5}{HDF5}{Hierarchical Data Format version 5}
\newacronym{hfss}{HFSS}{High Frequency Simulator Software}
\newacronym{hmd}{HMD}{head-mounted display}
\newacronym{hpbm}{HPBM}{half power beam width}
\newacronym{HyMPRo}{HyMPRo}{Hybrid Multi-Path Routing algorithm}
\newacronym{i.i.d.}{i.i.d.}{independent and identically distributed}
\newacronym{i2c}{I2C}{Inter-Integrated Circuit}
\newacronym{iaq}{IAQ}{Indoor Air Quality}
\newacronym{ib}{IB}{in-band}
\newacronym{ibo}{IBO}{input back-off}
\newacronym{ic}{IC}{integrated circuit}
\newacronym{ici}{ICI}{intercarrier interference}
\newacronym{icnirp}{ICNIRP}{International Commission on Non-Ionizing Radiation Protection}
\newacronym{id}{ID}{information decoding}
\newacronym{idft}{IDFT}{inverse discrete Fourier transform}
\newacronym{if}{IF}{intermediate-frequency}
\newacronym{iid}{i.i.d.}{independently and identically distributed}
\newacronym{iis}{IIS}{integrated information system}
\newacronym{im}{IM}{intermodulation}
\newacronym{imd}{IMD}{intermodulation distortion}
\newacronym{imu}{IMU}{inertial measurement unit}
\newacronym{io}{IO}{input/output}
\newacronym{ioe}{IoE}{Internet of Everything}
\newacronym{iot}{IoT}{Internet of Things}
\newacronym{ipt}{IPT}{inductive power transfer}
\newacronym{ipy}{IPY}{Interventions per Year}
\newacronym{iq}{IQ}{in-phase and quadrature}
\newacronym{iqi}{IQI}{IQ imbalance}
\newacronym{ir}{IR}{infrared}
\newacronym{isi}{ISI}{intersymbol interference}
\newacronym{ism}{ISM}{industrial, scientific and medical}
\newacronym{isp}{ISP}{internet service provider}
\newacronym{jesd}{JESD}{Joint Electron Devices Engineering Council}
\newacronym{jfet}{JFET}{junction field effect transistor}
\newacronym{kpi}{KPI}{key performance indicator}
\newacronym{kvi}{KVI}{key value indicator}
\newacronym{larva}{LARVA}{LARge Virtual Array}
\newacronym{lca}{LCA}{life cycle assessment}
\newacronym{lco}{LCO}{lithium cobalt oxide}
\newacronym{ldo}{LDO}{Low-dropout voltage regulator}
\newacronym{ldpc}{LDPC}{low-density parity-check}
\newacronym{led}{LED}{Light Emitting Diode}
\newacronym{less}{LESS}{Low Energy Scheduler Solution}
\newacronym{lfp}{LFP}{lithium iron phosphate}
\newacronym{lib}{LIB}{Lithium-Ion Battery}
\newacronym{lic}{LIC}{lithium-ion capacitor}
\newacronym{lid}{LID}{Lithium Iron Disulfide}
\newacronym{lidar}{LiDAR}{light detection and ranging}
\newacronym{liion}{Li-ion}{lithium-ion}
\newacronym{lipo}{LiPo}{lithium polymer}
\newacronym{lis}{LIS}{large intelligent surface}
\newacronym{llh}{LLH}{log-likelihood}
\newacronym{lmd}{LMD}{Lithium Manganese Dioxide}
\newacronym{lmmse}{LMMSE}{least minimum mean square error}
\newacronym{lmo}{LMO}{lithium ion manganese oxide}
\newacronym{lna}{LNA}{low-noise amplifier}
\newacronym{lo}{LO}{local oscillator}
\newacronym{lora}{LoRa}{long range}
\newacronym{lorawan}{LoRaWAN}{long-range wide-area network}
\newacronym{los}{LoS}{line-of-sight}
\newacronym{lp}{LP}{linear programming}
\newacronym{lpf}{LPF}{low-pass filter}
\newacronym{lpt}{LPT}{laser power transfer}
\newacronym{lpwa}{LPWA}{Low Power Wide Area}
\newacronym{lpwan}{LPWAN}{low-power wide-area network}
\newacronym{lpwans}{LPWANs}{Low-Power Wide-Area Networks}
\newacronym{lqi}{LQI}{link quality indicator}
\newacronym{lrelu}{LReLU}{leaky rectified linear unit}
\newacronym{lrt}{LRT}{likelihood-ratio test}
\newacronym{ls}{LS}{least squares}
\newacronym{lsa}{LSA}{large synthetic array}
\newacronym{lsf}{LSF}{large-scale fading}
\newacronym{lsfc}{LSFC}{large-scale fading component}
\newacronym{lstm}{LSTM}{Long Short-Term Memory}
\newacronym{ltc}{LTC}{lithium thionyl chloride}
\newacronym{lte}{LTE}{Long Term Evolution}
\newacronym{lti}{LTI}{linear time-invariant}
\newacronym{lto}{LTO}{lithium titanate}
\newacronym{lusta}{LUSTA 5G }{Logistique mUltimodale Sécuritaire Téléopérée \& Autonome 5G}
\newacronym{m2m}{M2M}{machine to machine}
\newacronym{mac}{MAC}{Medium Access Control}
\newacronym{mate}{MATE}{millimeter-wave MIMO testbed}
\newacronym{mc}{MC}{Monte Carlo}
\newacronym{mcl}{MCL}{Maximum Coupling Loss}
\newacronym{mcs}{MCS}{modulation and coding scheme}
\newacronym{mcu}{MCU}{microcontroller unit}
\newacronym{mec}{MEC}{multi-access edge computing}
\newacronym{mems}{MEMS}{micro-electromechanical systems}
\newacronym{mf}{MF}{matched filter}
\newacronym{mimo}{MIMO}{multiple-input multiple-output}
\newacronym{miso}{MISO}{multiple-input single-output}
\newacronym{ml}{ML}{machine learning}
\newacronym{mlp}{MLP}{multilayer perceptron}
\newacronym{mmic}{MMIC}{monolithic microwave integrated circuit}
\newacronym{mmimo}{mMIMO}{massive MIMO}
\newacronym{mmse}{MMSE}{minimum mean square error}
\newacronym{mmtc}{mMTC}{massive machine-typed communication}
\newacronym{mmwave}{mmWave}{millimeter wave}
\newacronym{mn}{MN}{matching network}
\newacronym{mosfet}{MOSFET}{metal-oxide semiconductor field effect transistor}
\newacronym{mpc}{MPC}{multipath component}
\newacronym{mppt}{MPPT}{maximum power point tracking}
\newacronym{mr}{MR}{maximum ratio}
\newacronym{mrc}{MRC}{maximum ratio combining}
\newacronym{mrc_em}{MRC}{maximum ratio combining}
\newacronym{mrc_EM}{MRC}{Magnetic Resonance Coupling}
\newacronym{mrt}{MRT}{maximum ratio transmission}
\newacronym{mse}{MSE}{mean square error}
\newacronym{msk}{MSK}{Minimum-Shift Keying}
\newacronym{mtc}{MTC}{Machine-Type Communication}
\newacronym{multi-rat}{Multi-RAT}{multiple radio access technology}
\newacronym{multirat}{Multi-RAT}{Multiple Radio Access Technology}
\newacronym{music}{MUSIC}{MUltiple SIgnal Classification}
\newacronym{navauwall}{NAVAUWALL}{AUtomated NAVigation in WALLonia}
\newacronym{nb}{NB}{narrowband}
\newacronym{nbiot}{NB-IoT}{narrowband IoT}
\newacronym{nca}{NCA}{nickel cobalt aluminum}
\newacronym{netcdf}{NetCDF}{Network Common Data Form}
\newacronym{nf}{NF}{noise figure}
\newacronym{nfc}{NFC}{near-field communication}
\newacronym{nfv}{NFV}{network function virtualization}
\newacronym{ngmn}{NGMN}{Next Generation Mobile Networks }
\newacronym{ni}{NI}{National Instruments}
\newacronym{nicd}{NiCd}{nikkel cadmium}
\newacronym{nimh}{NiMH}{nikkel metal hydride}
\newacronym{nlos}{NLoS}{non-line-of-sight}
\newacronym{nmc}{NMC}{nickel manganese cobalt}
\newacronym{nmos}{nMOS}{n-channel metal-oxide semiconductor}
\newacronym{nn}{NN}{neural network}
\newacronym{nnls}{NNLS}{non-negative least squares}
\newacronym{noma}{NOMA}{non-orthogonal multiple access}
\newacronym{np}{NP}{Neyman-Pearson}
\newacronym{npbch}{NPBCH}{Narrowband Physical Broadcast Channel}
\newacronym{npss}{NPSS}{Narrow Band Primay Synchronization Signal}
\newacronym{nr}{NR}{New Radio}
\newacronym{nrs}{NRS}{Narrow Band Reference Signal}
\newacronym{nsss}{NSSS}{Narrowband Secondary Synchronization Signal}
\newacronym{ntp}{NTP}{network time protocol}
\newacronym{oai}{OAI}{OpenAirInterface} 
\newacronym{obw}{OBW}{occupied bandwidth}
\newacronym{ofdm}{OFDM}{orthogonal frequency-division multiplexing}
\newacronym{ofdma}{OFDMA}{orthogonal frequency-division multiple access}
\newacronym{ofdmim}{OFDM-IM}{OFDM with index modulation}
\newacronym{oma}{OMA}{orthogonal multiple access}
\newacronym{oob}{OOB}{out-of-band}
\newacronym{ook}{OOK}{on-off keying}
\newacronym{oran}{O-RAN}{open radio-access network}
\newacronym{os}{OS}{operating system}
\newacronym{ota}{OTA}{over-the-air}
\newacronym{otaa}{OTAA}{over-the-air authentication}
\newacronym{p1}{P1}{Phase 1}
\newacronym{p2}{P2}{Phase 2}
\newacronym{p2p}{P2P}{point-to-point}
\newacronym{pa}{PA}{power amplifier}
\newacronym{pae}{PAE}{power-added efficiency}
\newacronym{pana}{PanA}{Panel A}
\newacronym{panb}{PanB}{Panel B}
\newacronym{papr}{PAPR}{peak-to-average power ratio}
\newacronym{pc}{PC}{pilot count}
\newacronym{pcb}{PCB}{printed circuit board}
\newacronym{pcg}{PCG}{power consumption gain}
\newacronym{pcie}{PCIe}{Peripheral Component Interconnect Express}
\newacronym{pcsi}{PCSI}{perfect channel state information}
\newacronym{pd}{PD}{powered device}
\newacronym{pdcch}{PDCCH}{physical downlink control channel}
\newacronym{pdf}{PDF}{probability density function}
\newacronym{pdp}{PDP}{power delay profile}
\newacronym{pdsch}{PDSCH}{physical downlink shared channel}
\newacronym{pe}{PE}{processing element}
\newacronym{peb}{PEB}{positioning error bound}
\newacronym{per}{PER}{packet error rate}
\newacronym{pet}{PET}{privacy enhancing technology}
\newacronym{pg}{PG}{path gain}
\newacronym{pgd}{PGD}{proximal gradient descent}
\newacronym{phy}{PHY}{physical}
\newacronym{pki}{PKI}{public key infrastucture}
\newacronym{pl}{PL}{path loss}
\newacronym{pll}{PLL}{phase-locked loop}
\newacronym{pmf}{PMF}{polymer microwave fiber}
\newacronym{pmu}{PMU}{Power Management Unit}
\newacronym{pn}{PN}{pseudo-noise}
\newacronym{poe}{PoE}{power-over-Ethernet}
\newacronym{pps}{1PPS}{pulse per second}
\newacronym{pr}{PR}{phase reversal}
\newacronym{prbs}{PRBs}{Physical Resource Blocks}
\newacronym{prs}{PRS}{Peripheral Reflex System}
\newacronym{ps}{PS}{Processing System}
\newacronym{psd}{PSD}{power spectral density}
\newacronym{pse}{PSE}{power sourcing equipment}
\newacronym{psk}{PSK}{phase shift keying}
\newacronym{psm}{PSM}{power saving mode}
\newacronym{pss}{PSS}{primary synchronisation signal}
\newacronym{ptp}{PTP}{precision-time protocol}
\newacronym{ptrs}{PTRS}{Phase-Tracking Reference Signals}
\newacronym{ptw}{PTW}{paging time window}
\newacronym{pv}{PV}{photovoltaic}
\newacronym{pw}{PW}{planar wavefront}
\newacronym{pwm}{PWM}{pulse width modulation}
\newacronym{qam}{QAM}{quadrature amplitude modulation}
\newacronym{qos}{QoS}{quality-of-service}
\newacronym{qpsk}{QPSK}{quadrature phase-shift keying}
\newacronym{qrrls}{QR-RLS}{QR decomposition based recursive least squares}
\newacronym{quadriga}{QuaDRiGa}{QUAsi Deterministic RadIo channel GenerAtor}
\newacronym{ra}{RA}{Random Access}
\newacronym{ram}{RAM}{random-access memory}
\newacronym{ran}{RAN}{radio access network}
\newacronym{rar}{RAR}{Random Access Response}
\newacronym{rat}{RAT}{radio access technology}
\newacronym{rb}{Rb}{Rubidium}
\newacronym{rbs}{RBS}{radio base station}
\newacronym{rbw}{RBW}{resolution bandwidth}
\newacronym{rcs}{RCS}{radar cross section}
\newacronym{rdl}{RDL}{redistribution layer}
\newacronym{re}{RE}{radio element}
\newacronym{relu}{ReLU}{rectified linear unit}
\newacronym{rf}{RF}{radio frequency}
\newacronym{rfeh}{RFEH}{radio frequency energy harvesting}
\newacronym{rfic}{RFIC}{radio-frequency integrated circuit}
\newacronym{rfid}{RFID}{radio frequency identification}
\newacronym{rfpt}{RFPT}{radio frequency power transfer}
\newacronym{rfsoc}{RFSoC}{Radio Frequency System-on-Chip}
\newacronym{rir}{RIR}{room impulse response}
\newacronym{ris}{RIS}{reflective intelligent surface}
\newacronym{rllmtc}{RLLMTC}{reliable low latency machine type communication}
\newacronym{rls}{RLS}{recursive least squares}
\newacronym{rms}{RMS}{root-mean-square}
\newacronym{rmse}{RMSE}{root-mean-square error}
\newacronym{rmt}{RMT}{random matrix theory}
\newacronym{rof}{RoF}{radio-over-fiber}
\newacronym{ros}{ROS}{robot operating system}
\newacronym{rpi}{RPi}{Raspberry Pi}
\newacronym{rrc}{RRC}{Radio Resource Connection}
\newacronym{rreq}{RREQ}{route request packet}
\newacronym{rsrp}{RSRP}{Reference Signals Received Power}
\newacronym{rsrq}{RSRQ}{Reference Signal Received Quality}
\newacronym{rss}{RSS}{received signal strength}
\newacronym{rssi}{RSSI}{received signal strength indicator}
\newacronym{rtc}{RTC}{real time clock}
\newacronym{rtf}{RTF}{reader talks first}
\newacronym{rtk}{RTK}{real time kinematics}
\newacronym{rts}{RTS}{ray tracing simulator}
\newacronym{ru}{RU}{Radio Unit}
\newacronym{rv}{RV}{random variable}
\newacronym{rw}{RW}{RadioWeaves}
\newacronym{rx}{RX}{receiver}
\newacronym{rzf}{RZF}{regularized zero forcing}
\newacronym{s-parameter}{S-parameter}{scattering parameter}
\newacronym{sa}{SA}{synchronization anchor}
\newacronym{sa5}{SA}{Stand Alone}
\newacronym{sar}{SAR}{specific absorption rate}
\newacronym{sbl}{SBL}{sparse Bayesian learning}
\newacronym{scfdma}{SCFDMA}{single-carrier frequency division multiple access}
\newacronym{sdg}{SDG}{Sustainable Development Goal}
\newacronym{sdm}{SDM}{sigma-delta modulator}
\newacronym{sdma}{SDMA}{spatial-division multiple access}
\newacronym{sdn}{SDN}{software-defined network}
\newacronym{sdof}{SDoF}{sigma-delta over fiber}
\newacronym{sdr}{SDR}{software-defined radio}
\newacronym{se}{SE}{spectral efficiency}
\newacronym{sei}{SEI}{specific emitter identification}
\newacronym{ser}{SER}{symbol-error rate}
\newacronym{sf}{SF}{spreading factor}
\newacronym{sfn}{SFN}{single frequency network}
\newacronym{sfp}{SFP}{small form-factor pluggable}
\newacronym{sha}{SHA}{Secure Hash Algorithm}
\newacronym{SigMF}{SigMF}{Signal Metadata Format}
\newacronym{simo}{SIMO}{single-input multiple-output}
\newacronym{sinr}{SINR}{signal-to-interference-plus-noise ratio}
\newacronym{siso}{SISO}{single-input single-output}
\newacronym{slam}{SLAM}{simultaneous localization and mapping}
\newacronym{slc}{SLC}{spatial leakage suppression}
\newacronym{slerp}{SLERP}{spherical linear interpolation}
\newacronym{sma}{SMA}{SubMiniature version A}
\newacronym{smc}{SMC}{specular multipath component}
\newacronym{smps}{SMPS}{switched mode power supply}
\newacronym{sndr}{SNDR}{signal-to-noise-and-distortion ratio}
\newacronym{snidr}{SNIDR}{signal-to-noise-and-interference-and-distortion ratio}
\newacronym{snir}{SNIR}{signal-to-interference-plus-noise ratio}
\newacronym{snr}{SNR}{signal-to-noise ratio}
\newacronym{soc}{SoC}{state of charge}
\newacronym{SoC}{SOC}{System on Chip}
\newacronym{sp}{SP}{service point}
\newacronym{spdt}{SPDT}{single pole double throw}
\newacronym{spi}{SPI}{Serial Peripheral Interface}
\newacronym{spst}{SPST}{single pole single throw}
\newacronym{sram}{SRAM}{static random-access memory}
\newacronym{srd}{SRD}{short-range device}
\newacronym{srls}{SRLS}{standard recursive least squares}
\newacronym{srs}{SRS}{Sounding Reference Signal}
\newacronym{ssb}{SSB}{synchronisation signal block}
\newacronym{ssd}{SSD}{solid state drive}
\newacronym{ssq}{SSQ}{simulator sickness questionnaire}
\newacronym{steam}{STEAM}{science, technology, engineering, the arts, and mathematics}
\newacronym{svd}{SVD}{singular value decomposition}
\newacronym{sw}{SW}{spherical wavefront}
\newacronym{swipt}{SWIPT}{simultaneous wireless information and power transfer}
\newacronym{synce}{SyncE}{Synchronous Ethernet}
\newacronym{tau}{TAU}{tracking area update}
\newacronym{tcer}{TCER}{transported to consumed energy ratio}
\newacronym{tcp}{TCP}{Transmission Control Protocol}
\newacronym{tcxo}{TCXO}{temperature compensated crystal oscillator}
\newacronym{tdd}{TDD}{time division duplexing}
\newacronym{tdma}{TDMA}{time division-multiple access}
\newacronym{tdoa}{TDOA}{time-difference-of-arrival}
\newacronym{thz}{THz}{Terahertz}
\newacronym{to}{TO}{timing offset}
\newacronym{toa}{TOA}{time-of-arrival}
\newacronym{tof}{ToF}{time-of-flight}
\newacronym{tosm}{TOSM}{through-open-short-match}
\newacronym{tpms}{TPMS}{Tire-Pressure Monitoring System}
\newacronym{trl}{TRL}{technology readyness level}
\newacronym{trp}{TRP}{Transmission Reception Point}
\newacronym{tsn}{TSN}{time-sensitive networking}
\newacronym{ttf}{TTF}{tag talks first}
\newacronym{ttff}{TTFF}{Time To First Fix}
\newacronym{ttm}{TTM}{time to market}
\newacronym{ttn}{TTN}{The Things Network}
\newacronym{tx}{TX}{transmitter}
\newacronym{uart}{UART}{Universal Asynchronous Receiver/Transmitter}
\newacronym{uav}{UAV}{unmanned aerial vehicle}
\newacronym{uc}{UC}{use case}
\newacronym{ucie}{UCIe}{Universal Chiplet Interconnect Express}
\newacronym{udp}{UDP}{User Datagram Protocol}
\newacronym{ue}{UE}{user equipment}
\newacronym{ugv}{UGV}{unmanned ground vehicle}
\newacronym{uhd}{UHD}{USRP hardware driver}
\newacronym{uhf}{UHF}{ultra-high frequency}
\newacronym{ul}{UL}{uplink}
\newacronym{ula}{ULA}{uniform linear array}
\newacronym{uMIMO}{$\mu$-MIMO}{ultra-massive Multiple-Input Multiple-Output}
\newacronym{ummtc}{umMTC}{ultra massive machine type communication}
\newacronym{un}{UN}{United Nations}
\newacronym{upa}{UPA}{uniform planar array}
\newacronym{ura}{URA}{uniform rectangular array}
\newacronym{urllc}{URLLC}{ultra-reliable low-latency communications}
\newacronym{usrp}{USRP}{universal software radio peripheral}
\newacronym{uv}{UV}{unmanned vehicle}
\newacronym{uwb}{UWB}{ultrawideband}
\newacronym{v2v}{V2V}{vehicle-to-vehicle}
\newacronym{vco}{VCO}{voltage-controlled oscillator}
\newacronym{vep}{VEP}{virtual edge platform}
\newacronym{vlc}{VLC}{visible light communication}
\newacronym{vlp}{VLP}{visible light positioning}
\newacronym{vna}{VNA}{vector network analyzer}
\newacronym{voc}{VOC}{Voltatile Organic Compound}
\newacronym{votable}{VOTable}{Virtual Observatory Table}
\newacronym{vr}{VR}{virtual reality}
\newacronym{wb}{WB}{wideband}
\newacronym{wban}{WBAN}{wireless body area network}
\newacronym{wd}{WD}{Wireless distance}
\newacronym{wimax}{WiMAX}{Worldwide Interoperability for Microwave Access}
\newacronym{wlan}{WLAN}{wireless LAN}
\newacronym{wpt}{WPT}{wireless power transfer}
\newacronym{wr}{WR}{White Rabbit}
\newacronym{wrsn}{WRSN}{wireless rechargeable sensor network}
\newacronym{wsn}{WSN}{wireless sensor network}
\newacronym{xets}{XETS}{cross exponentially tapered slot}
\newacronym{xr}{XR}{extended reality}
\newacronym{z3ro}{Z3RO}{zero third-order distortion}
\newacronym{zf}{ZF}{zero-forcing}
\newacronym{zmcscg}{ZMCSCG}{zero mean circularly symmetric complex Gaussian}
\newacronym{ep}{EP}{energy profiler}

%% file: auto_names.tex
\makeatletter
\PassOptionsToPackage{dvipsnames,svgnames,table}{xcolor}%
\makeatother

\usepackage{listofitems}
\usepackage{pgffor}
\usepackage{xparse}
\usepackage{ifthen}
\usepackage{tikz}
\usepackage{xspace}
\usetikzlibrary{external}

\makeatletter
\def\myColorList{LimeGreen,BrickRed,Fuchsia,Bittersweet,YellowOrange, YellowGreen, WildStrawberry, Fuchsia, RedViolet, Periwinkle, PineGreen, OrangeRed, RawSienna, Turquoise, Aquamarine, BlueViolet, BurntOrange, DarkOrchid, TealBlue, Violet, RoyalPurple}
\readlist\ColorList\myColorList

\definecolor{defaultColor}{RGB}{112,45,128}
\NewDocumentCommand{\roundLabel}{O{defaultColor} m}{%
\tikzexternaldisable
    \tikz[baseline=(char.base)]{
        \node[rectangle, rounded corners=0.65mm, inner sep=0.65mm, fill=#1, draw=#1, text=white, font=\itshape](char) {#2};
    }%
\tikzexternalenable
}

\NewDocumentCommand{\defineauthors}{ O{true} m }{%
    \ifthenelse{\equal{#1}{true}}{%
        Personal commands for comments are:
        \begin{itemize}
        \foreach \x [count=\xi from 1] in {#2} {   
            \item \textcolor{\ColorList[\xi]}{\textbackslash\unexpanded\expandafter{\x}\{\}}
            }
        \end{itemize}
        \foreach \x [count=\xi from 1] in {#2} { 
            \expandafter\xdef\csname\x\endcsname####1{\noexpand\textcolor{\ColorList[\xi]}{\roundLabel[\ColorList[\xi]]{\unexpanded\expandafter{\x}} ####1}}%
        }
    }{%
        \foreach \x [count=\xi from 1] in {#2} { 
            \expandafter\xdef\csname\x\endcsname####1{\noexpand\textcolor{\ColorList[\xi]}{\roundLabel[\ColorList[\xi]]{\unexpanded\expandafter{\x}} ####1}}%
        }
    }
    \foreach \x [count=\xi from 1] in {#2} {
        \expandafter\xdef\csname at\x\endcsname####1{\noexpand\textcolor{\ColorList[\xi]}{@\x}}%
    }
}

\makeatother

%% file: tikz/harv-efficiency.tex
\begin{tikzpicture}

\definecolor{darkgray176}{RGB}{176,176,176}
\definecolor{darkorange25512714}{RGB}{255,127,14}
\definecolor{lightgray204}{RGB}{204,204,204}
\definecolor{steelblue31119180}{RGB}{31,119,180}
\definecolor{green441644}{RGB}{44,160,44}

\begin{axis}[
legend cell align={left},
legend style={
  fill opacity=0.8,
  draw opacity=1,
  text opacity=1,
  at={(0.03,0.97)},
  anchor=north west,
  draw=lightgray204
},
width=0.85\textwidth,
height=0.28\textwidth,
tick align=outside,
tick pos=left,
x grid style={darkgray176},
xmajorgrids,
xlabel={Total radiated power of $84$ antennas combined [\si{\dBm}]},
xmin=28, xmax=37,
xtick style={color=black},
ylabel={Power [uW]},
ymin=0, ymax=26,
ytick style={color=black}
]
\addplot [semithick, steelblue31119180]
table {%
28.34 0.563309137141809
29.2  0.617264728440383
30.06 1.19676419772738
30.92 1.62401733676224
31.78 2.42358335144312
32.64 3.41778224120492
33.44 4.62544215112466
34.24 6.1594953739722
35.04 6.88141549103359
35.84 10.2900477634388
36.64 12.5641747503154
};
\addlegendentry{\small {\color{steelblue31119180} Single-tone mean DC power} \measOne{}}
\addplot [semithick, black]
table {%
28.34 3.3811171551552
29.2  3.832819324565
30.06 5.17649837949025
30.92 6.48721963079147
31.78 7.85706322014636
32.64 10.5238226401887
33.44 12.4856654643546
34.24 15.1302972269924
35.04 16.5619840572225
35.84 21.0765063168423
36.64 24.6765328148914
};
\addlegendentry{\small Single-tone mean RF power \measOne{}}
\addplot [semithick, steelblue31119180, densely dashed]
table {%
28.34 0.456691991671377
29.2  0.63950951171412
30.06 0.921566281708466
30.92 1.26602254753809
31.78 1.66476190894102
32.64 2.14176685947878
33.44 2.63355270233796
34.24 3.10860562078332
35.04 3.8525379597825
35.84 6.26595860409853
36.64 8.04211120327822
};
\addlegendentry{\small {\color{steelblue31119180} Multi-tone mean DC power \measTwo{}}}
\addplot [semithick, black, densely dashed]
table {%
28.34 3.1382763220479
29.2  3.84886589919538
30.06 4.81378016660812
30.92 6.04560337368143
31.78 7.3304940846566
32.64 9.38295576066426
33.44 11.3880508544913
34.24 13.9698884109967
35.04 17.0795044737649
35.84 21.0162069983013
36.64 25.0036933829554
};
\addlegendentry{\small Multi-tone mean RF power \measTwo{}}
\end{axis}

\begin{axis}[
width=0.85\textwidth,
height=0.28\textwidth,
legend cell align={center},
legend style={fill opacity=0.8, draw opacity=1, text opacity=1, at={(0.6,0.97)}, anchor=north, draw=lightgray204},
tick align=outside,
xmin=28, xmax=37,
xtick pos=left,
xtick style={color=black},
xticklabels=\empty, 
ylabel=\textcolor{darkorange25512714}{Efficiency [\%]},
grid=none,
ymin=0, ymax=52,
ytick pos=right,
ytick style={color=black},
yticklabel style={anchor=west},
axis line style={-}
]
\addplot [semithick, darkorange25512714]
table {%
28.34 16.6604442050442
29.2  16.1047176026342
30.06 23.1191842437171
30.92 25.034104426708
31.78 30.8459189335373
32.64 32.4766233531244
33.44 37.0460202087734
34.24 40.7096786108317
35.04 41.5494633206863
35.84 48.8223598766744
36.64 50.9154784611124
};
\addlegendentry{\small Single-tone efficiency \measOne{}}
\addplot [semithick, darkorange25512714, densely dashed]
table {%
28.34 14.5523193245571
29.2  16.6155311321138
30.06 19.1443366712323
30.92 20.9412108152764
31.78 22.7100914306107
32.64 22.8261425728725
33.44 23.1255790476148
34.24 22.2521864837253
35.04 22.5564972666521
35.84 29.8148881223192
36.64 32.1636930996778
};
\addlegendentry{\small Multi-tone efficiency \measTwo{}}
\end{axis}

\end{tikzpicture}

%% file: tikz/one_tone_phase_duration_5_m1_80_energy_buffer.tex
\begin{tikzpicture}

\definecolor{darkgray176}{RGB}{176,176,176}
\definecolor{darkorange25512714}{RGB}{255,127,14}
\definecolor{lightgray204}{RGB}{204,204,204}
\definecolor{steelblue31119180}{RGB}{31,119,180}

\begin{axis}[
width=\columnwidth,
height=0.55\columnwidth,
line cap = round,
legend cell align={left},
legend style={
  fill opacity=0.8,
  draw opacity=1,
  text opacity=1,
  at={(0.97,0.03)},
  anchor=south east,
  draw=lightgray204
},
tick align=outside,
tick pos=left,
x grid style={darkgray176},
xlabel={Time [s]},
xmajorgrids,
xmin=-9.99475, xmax=209.95575,
xtick style={color=black},
y grid style={darkgray176},
ylabel={Voltage [V]},
ymajorgrids,
ymin=-0.0746080000000081, ymax=2.1,
ytick style={color=black}
]
\addplot [semithick, steelblue31119180]
table {%
0.00300000000000011 1.704
0.203999999999994 1.701
0.406000000000006 1.706
0.606999999999999 1.707
0.807999999999993 1.708
1.01000000000001 1.709
1.212 1.707
1.41200000000001 1.71
1.61499999999999 1.707
1.816 1.71
2.017 1.706
2.218 1.702
2.42100000000001 1.699
2.621 1.701
2.82299999999999 1.702
3.024 1.704
3.226 1.706
3.42699999999999 1.708
3.628 1.712
3.83 1.707
4.032 1.709
4.232 0.621
4.434 0.623
4.63500000000001 0.618
4.837 0.62
5.038 0.621
5.23999999999999 0.619
5.441 0.615
5.643 0.613
5.84399999999999 0.61
6.04600000000001 0.609
6.246 0.613
6.44799999999999 0.612
6.65000000000001 0.617
6.851 0.621
7.053 0.624
7.254 0.621
7.455 0.616
7.657 0.615
7.858 0.612
8.06 0.614
8.261 0.612
8.46299999999999 0.606
8.664 0.607
8.866 0.61
9.06699999999999 0.613
9.26900000000001 0.465
9.46899999999999 0.46
9.67100000000001 0.459
9.873 0.458
10.074 0.46
10.275 0.46
10.476 0.454
10.678 0.449
10.88 0.45
11.081 0.454
11.283 0.456
11.484 0.459
11.685 0.461
11.886 0.464
12.088 0.456
12.289 0.462
12.491 0.456
12.692 0.456
12.894 0.455
13.095 0.465
13.296 0.449
13.498 0.452
13.699 0.452
13.901 0.455
14.102 1.308
14.304 1.305
14.505 1.311
14.706 1.312
14.908 1.314
15.109 1.312
15.311 1.314
15.512 1.313
15.714 1.308
15.915 1.31
16.117 1.301
16.317 1.302
16.519 1.296
16.72 1.303
16.923 1.312
17.123 1.31
17.325 1.311
17.526 1.313
17.728 1.314
17.929 1.313
18.13 1.309
18.333 1.308
18.534 1.307
18.734 1.308
18.936 1.307
19.138 0.777
19.339 0.759
19.54 0.765
19.742 0.76
19.943 0.764
20.144 0.758
20.345 0.767
20.548 0.771
20.749 0.772
20.951 0.786
21.151 0.78
21.353 0.789
21.554 0.782
21.756 0.779
21.958 0.772
22.159 0.761
22.36 0.762
22.561 0.762
22.763 0.767
22.965 0.764
23.166 0.773
23.367 0.772
23.569 0.776
23.771 0.775
23.972 0.785
24.173 1.85
24.374 1.848
24.576 1.844
24.777 1.836
24.979 1.835
25.18 1.837
25.38 1.838
25.583 1.841
25.784 1.841
25.985 1.845
26.187 1.843
26.388 1.844
26.59 1.84
26.791 1.844
26.993 1.843
27.193 1.846
27.394 1.845
27.596 1.84
27.797 1.835
27.999 1.831
28.2 1.836
28.402 1.836
28.603 1.84
28.805 1.839
29.006 1.844
29.208 0.788
29.408 0.787
29.61 0.791
29.812 0.79
30.013 0.789
30.215 0.787
30.416 0.792
30.617 0.791
30.819 0.778
31.021 0.78
31.222 0.767
31.423 0.776
31.625 0.78
31.866 0.801
32.068 0.787
32.269 0.793
32.471 0.792
32.672 0.79
32.873 0.793
33.075 0.792
33.276 0.795
33.478 0.786
33.678 0.783
33.881 0.782
34.083 0.495
34.283 0.495
34.485 0.495
34.686 0.5
34.887 0.499
35.09 0.5
35.29 0.499
35.492 0.495
35.693 0.493
35.895 0.493
36.096 0.494
36.297 0.498
36.499 0.486
36.7 0.489
36.902 0.492
37.103 0.496
37.305 0.496
37.506 0.495
37.708 0.5
37.909 0.499
38.11 0.499
38.312 0.498
38.513 0.493
38.715 0.492
38.915 0.495
39.117 1.798
39.319 1.794
39.52 1.794
39.721 1.789
39.923 1.788
40.125 1.792
40.325 1.792
40.526 1.794
40.728 1.793
40.929 1.798
41.132 1.798
41.332 1.798
41.534 1.797
41.735 1.798
41.937 1.795
42.138 1.793
42.34 1.794
42.541 1.788
42.743 1.787
42.944 1.787
43.146 1.793
43.347 1.794
43.548 1.795
43.749 1.794
43.951 1.797
44.153 1.523
44.354 1.522
44.556 1.524
44.757 1.524
44.959 1.522
45.16 1.525
45.362 1.527
45.563 1.513
45.764 1.515
45.966 1.513
46.167 1.514
46.368 1.517
46.57 1.519
46.771 1.519
46.973 1.519
47.174 1.519
47.375 1.519
47.576 1.522
47.779 1.521
47.979 1.523
48.182 1.527
48.382 1.528
48.584 1.519
48.785 1.512
48.986 1.51
49.188 1.601
49.389 1.602
49.591 1.606
49.792 1.605
49.994 1.61
50.195 1.606
50.397 1.604
50.597 1.605
50.799 1.6
51 1.602
51.202 1.599
51.403 1.606
51.605 1.597
51.806 1.594
52.008 1.596
52.208 1.599
52.411 1.604
52.612 1.601
52.813 1.607
53.014 1.606
53.216 1.609
53.418 1.603
53.618 1.605
53.82 1.602
54.022 1.604
54.223 1.504
54.425 1.505
54.626 1.501
54.828 1.504
55.029 1.502
55.231 1.503
55.432 1.502
55.633 1.504
55.835 1.497
56.036 1.507
56.238 1.505
56.438 1.508
56.641 1.512
56.842 1.511
57.043 1.51
57.245 1.506
57.445 1.503
57.647 1.503
57.848 1.499
58.051 1.499
58.251 1.502
58.453 1.497
58.654 1.502
58.856 1.501
59.057 1.505
59.259 0.482
59.459 0.481
59.661 0.479
59.862 0.478
60.064 0.477
60.266 0.479
60.466 0.477
60.668 0.47
60.87 0.472
61.071 0.479
61.272 0.48
61.475 0.477
61.675 0.475
61.877 0.482
62.079 0.48
62.28 0.481
62.481 0.482
62.682 0.476
62.884 0.478
63.085 0.479
63.287 0.477
63.488 0.474
63.689 0.471
63.89 0.468
64.093 1.185
64.293 1.187
64.495 1.185
64.696 1.187
64.898 1.185
65.099 1.194
65.301 1.189
65.502 1.194
65.704 1.194
65.905 1.197
66.106 1.196
66.308 1.184
66.51 1.193
66.711 1.188
66.912 1.19
67.114 1.186
67.315 1.186
67.517 1.187
67.717 1.188
67.92 1.191
68.121 1.19
68.322 1.194
68.524 1.19
68.724 1.196
68.927 1.194
69.127 1.843
69.33 1.846
69.531 1.838
69.731 1.836
69.932 1.828
70.135 1.828
70.335 1.834
70.537 1.834
70.738 1.838
70.94 1.84
71.142 1.842
71.342 1.839
71.545 1.838
71.745 1.838
71.947 1.84
72.148 1.843
72.35 1.841
72.551 1.844
72.753 1.832
72.954 1.831
73.155 1.827
73.358 1.833
73.558 1.833
73.76 1.839
73.961 1.846
74.163 0.699
74.364 0.699
74.565 0.704
74.768 0.705
74.969 0.701
75.17 0.699
75.371 0.688
75.572 0.695
75.775 0.684
75.975 0.688
76.177 0.687
76.378 0.689
76.58 0.689
76.781 0.69
76.981 0.692
77.184 0.695
77.385 0.7
77.587 0.704
77.787 0.705
77.989 0.702
78.19 0.695
78.392 0.696
78.593 0.688
78.795 0.693
78.996 0.687
79.198 1.838
79.399 1.841
79.6 1.841
79.801 1.841
80.002 1.845
80.204 1.843
80.406 1.847
80.607 1.844
80.808 1.846
81.01 1.846
81.212 1.844
81.412 1.843
81.615 1.844
81.816 1.838
82.017 1.834
82.219 1.834
82.419 1.836
82.621 1.842
82.822 1.839
83.024 1.84
83.225 1.845
83.427 1.844
83.628 1.846
83.829 1.845
84.063 1.839
84.264 1.573
84.466 1.571
84.668 1.576
84.869 1.576
85.07 1.575
85.272 1.573
85.473 1.572
85.675 1.573
85.875 1.575
86.078 1.569
86.279 1.566
86.479 1.569
86.681 1.57
86.882 1.572
87.085 1.57
87.285 1.571
87.487 1.577
87.689 1.576
87.89 1.573
88.091 1.574
88.293 1.574
88.494 1.575
88.696 1.576
88.897 1.575
89.099 0.069
89.3 0.068
89.502 0.068
89.703 0.069
89.904 0.068
90.106 0.069
90.307 0.069
90.508 0.069
90.71 0.069
90.912 0.069
91.113 0.069
91.314 0.069
91.515 0.068
91.717 0.068
91.919 0.068
92.119 0.068
92.321 0.068
92.522 0.068
92.724 0.068
92.925 0.068
93.126 0.069
93.328 0.069
93.529 0.069
93.731 0.068
93.932 0.068
94.133 0.48
94.335 0.481
94.536 0.479
94.738 0.474
94.939 0.475
95.141 0.48
95.342 0.477
95.544 0.477
95.744 0.474
95.947 0.475
96.148 0.476
96.349 0.476
96.551 0.478
96.751 0.482
96.954 0.483
97.154 0.482
97.357 0.48
97.557 0.484
97.76 0.471
97.961 0.465
98.162 0.478
98.364 0.472
98.564 0.475
98.766 0.473
98.968 0.473
99.169 0.672
99.371 0.676
99.572 0.68
99.773 0.673
99.975 0.669
100.177 0.663
100.377 0.664
100.58 0.659
100.78 0.679
100.982 0.665
101.184 0.652
101.385 0.662
101.587 0.67
101.787 0.679
101.989 0.668
102.19 0.672
102.392 0.67
102.593 0.669
102.795 0.668
102.995 0.663
103.198 0.668
103.399 0.661
103.6 0.667
103.802 0.657
104.003 0.657
104.204 0.051
104.406 0.051
104.607 0.051
104.809 0.051
105.01 0.051
105.211 0.05
105.413 0.05
105.614 0.05
105.816 0.05
106.017 0.05
106.219 0.05
106.42 0.051
106.622 0.051
106.823 0.051
107.025 0.05
107.226 0.05
107.428 0.051
107.629 0.051
107.831 0.05
108.032 0.05
108.233 0.05
108.434 0.05
108.636 0.05
108.837 0.051
109.038 0.051
109.24 1.836
109.441 1.835
109.643 1.831
109.844 1.837
110.046 1.839
110.247 1.841
110.449 1.846
110.65 1.844
110.851 1.844
111.052 1.84
111.254 1.839
111.455 1.839
111.657 1.841
111.858 1.839
112.059 1.835
112.261 1.835
112.462 1.832
112.664 1.84
112.865 1.839
113.067 1.845
113.268 1.845
113.47 1.843
113.671 1.842
113.872 1.841
114.073 1.84
114.276 1.315
114.477 1.317
114.678 1.314
114.88 1.311
115.081 1.312
115.282 1.314
115.483 1.315
115.685 1.314
115.886 1.314
116.088 1.321
116.29 1.323
116.492 1.32
116.692 1.321
116.894 1.316
117.096 1.314
117.297 1.314
117.498 1.309
117.698 1.308
117.9 1.31
118.103 1.313
118.303 1.313
118.505 1.313
118.706 1.317
118.909 1.322
119.11 0.06
119.31 0.06
119.512 0.06
119.713 0.06
119.915 0.061
120.117 0.06
120.318 0.061
120.519 0.061
120.721 0.06
120.922 0.06
121.123 0.061
121.325 0.061
121.526 0.061
121.727 0.061
121.929 0.06
122.13 0.06
122.332 0.06
122.534 0.061
122.735 0.061
122.937 0.06
123.138 0.061
123.34 0.06
123.541 0.06
123.742 0.061
123.943 0.062
124.144 1.051
124.346 1.054
124.547 1.053
124.749 1.045
124.95 1.041
125.151 1.042
125.354 1.051
125.554 1.038
125.756 1.03
125.958 1.034
126.159 1.043
126.361 1.045
126.562 1.045
126.763 1.052
126.964 1.05
127.166 1.053
127.368 1.044
127.569 1.046
127.77 1.046
127.971 1.047
128.172 1.053
128.375 1.037
128.576 1.034
128.778 1.039
128.978 1.047
129.18 1.469
129.38 1.471
129.582 1.473
129.784 1.475
129.985 1.477
130.186 1.478
130.389 1.48
130.589 1.475
130.791 1.475
130.993 1.47
131.194 1.467
131.395 1.467
131.597 1.468
131.797 1.472
131.999 1.47
132.201 1.47
132.403 1.474
132.603 1.478
132.806 1.475
133.007 1.475
133.207 1.476
133.409 1.475
133.61 1.473
133.812 1.474
134.013 1.469
134.215 1.814
134.415 1.813
134.617 1.816
134.818 1.816
135.02 1.817
135.222 1.821
135.423 1.819
135.625 1.82
135.825 1.819
136.027 1.819
136.229 1.816
136.43 1.817
136.632 1.818
136.833 1.813
137.034 1.815
137.235 1.814
137.438 1.817
137.638 1.817
137.84 1.817
138.041 1.818
138.243 1.82
138.444 1.819
138.645 1.817
138.846 1.817
139.047 1.816
139.249 1.116
139.451 1.116
139.652 1.116
139.853 1.115
140.096 1.109
140.297 1.11
140.499 1.107
140.7 1.109
140.9 1.113
141.102 1.116
141.303 1.116
141.505 1.116
141.706 1.126
141.908 1.123
142.109 1.121
142.311 1.113
142.512 1.114
142.713 1.109
142.915 1.111
143.116 1.107
143.318 1.106
143.519 1.111
143.721 1.11
143.922 1.112
144.123 1.819
144.325 1.818
144.525 1.819
144.727 1.82
144.929 1.82
145.13 1.816
145.331 1.818
145.533 1.818
145.734 1.817
145.936 1.816
146.138 1.818
146.338 1.818
146.54 1.818
146.741 1.818
146.943 1.819
147.144 1.819
147.346 1.817
147.547 1.82
147.748 1.82
147.95 1.821
148.151 1.815
148.353 1.817
148.553 1.817
148.755 1.815
148.956 1.816
149.158 1.784
149.359 1.787
149.561 1.788
149.762 1.787
149.964 1.787
150.165 1.789
150.366 1.789
150.567 1.789
150.77 1.789
150.971 1.787
151.173 1.785
151.373 1.784
151.574 1.786
151.776 1.784
151.977 1.787
152.179 1.786
152.38 1.79
152.582 1.789
152.783 1.788
152.985 1.791
153.186 1.79
153.387 1.79
153.589 1.79
153.789 1.79
153.991 1.786
154.193 1.559
154.394 1.556
154.595 1.554
154.798 1.562
154.998 1.56
155.2 1.559
155.401 1.557
155.602 1.564
155.804 1.56
156.005 1.561
156.207 1.564
156.409 1.568
156.61 1.567
156.812 1.565
157.013 1.564
157.215 1.558
157.415 1.562
157.617 1.564
157.818 1.564
158.02 1.563
158.221 1.559
158.422 1.562
158.624 1.561
158.824 1.561
159.027 1.565
159.227 0.958
159.429 0.959
159.63 0.963
159.833 0.95
160.034 0.941
160.235 0.943
160.436 0.944
160.638 0.946
160.839 0.95
161.041 0.959
161.242 0.958
161.444 0.956
161.645 0.953
161.846 0.954
162.048 0.956
162.249 0.957
162.45 0.961
162.652 0.954
162.854 0.946
163.054 0.942
163.256 0.946
163.458 0.948
163.659 0.952
163.86 0.959
164.062 0.964
164.263 0.194
164.465 0.194
164.665 0.19
164.867 0.19
165.069 0.19
165.271 0.189
165.471 0.19
165.673 0.191
165.874 0.188
166.076 0.188
166.277 0.19
166.479 0.189
166.68 0.188
166.882 0.187
167.082 0.194
167.283 0.194
167.486 0.189
167.687 0.191
167.889 0.189
168.09 0.185
168.291 0.187
168.492 0.186
168.694 0.185
168.895 0.186
169.097 0.538
169.299 0.53
169.5 0.531
169.702 0.535
169.902 0.538
170.104 0.54
170.306 0.539
170.507 0.538
170.708 0.536
170.909 0.534
171.111 0.535
171.313 0.532
171.514 0.53
171.715 0.532
171.916 0.541
172.118 0.533
172.32 0.534
172.52 0.536
172.723 0.541
172.923 0.541
173.125 0.541
173.327 0.535
173.528 0.536
173.729 0.534
173.931 0.534
174.132 0.375
174.334 0.369
174.535 0.369
174.736 0.368
174.938 0.365
175.139 0.367
175.34 0.371
175.542 0.375
175.744 0.377
175.944 0.38
176.147 0.38
176.347 0.376
176.549 0.375
176.75 0.372
176.952 0.372
177.153 0.369
177.354 0.364
177.556 0.366
177.757 0.368
177.959 0.364
178.16 0.371
178.362 0.375
178.563 0.377
178.764 0.375
178.966 0.379
179.167 1.846
179.368 1.845
179.57 1.839
179.771 1.836
179.973 1.842
180.174 1.842
180.375 1.844
180.577 1.841
180.778 1.844
180.98 1.842
181.181 1.847
181.383 1.844
181.584 1.845
181.786 1.845
181.987 1.838
182.189 1.844
182.39 1.841
182.592 1.841
182.792 1.838
182.993 1.844
183.195 1.843
183.396 1.844
183.598 1.842
183.799 1.843
184.001 1.845
184.202 1.843
184.404 1.843
184.605 1.844
184.807 1.838
185.008 1.838
185.21 1.828
185.411 1.827
185.613 1.827
185.814 1.828
186.016 1.83
186.216 1.836
186.418 1.834
186.619 1.838
186.821 1.834
187.022 1.836
187.224 1.834
187.425 1.834
187.627 1.839
187.828 1.838
188.03 1.83
188.231 1.827
188.432 1.83
188.634 1.83
188.835 1.833
189.037 1.834
189.238 1.655
189.44 1.653
189.641 1.654
189.842 1.653
190.044 1.651
190.245 1.65
190.447 1.649
190.648 1.651
190.85 1.651
191.051 1.644
191.293 1.642
191.494 1.646
191.696 1.644
191.897 1.652
192.098 1.65
192.299 1.656
192.501 1.653
192.703 1.653
192.904 1.652
193.106 1.649
193.307 1.65
193.508 1.651
193.71 1.645
193.911 1.643
194.113 0.478
194.314 0.478
194.516 0.479
194.717 0.479
194.918 0.48
195.12 0.482
195.32 0.481
195.522 0.483
195.724 0.485
195.926 0.481
196.127 0.483
196.329 0.483
196.529 0.482
196.731 0.479
196.932 0.477
197.134 0.48
197.336 0.477
197.537 0.478
197.739 0.479
197.939 0.481
198.142 0.481
198.342 0.483
198.545 0.484
198.745 0.484
198.947 0.484
199.148 1.181
199.349 1.173
199.551 1.171
199.752 1.167
199.954 1.181
};
\addlegendentry{\small Harvester voltage}
\addplot [semithick, darkorange25512714]
table {%
0.00799999999999557 0.0170399999999923
0.207999999999998 0.109168114111796
0.409999999999997 0.153831002064275
0.611000000000004 0.188025468085657
0.811999999999998 0.216886117208864
1.014 0.242476957899879
1.21599999999999 0.265587507802531
1.416 0.286670612727613
1.619 0.306583359854698
1.81999999999999 0.325102056360541
2.021 0.342594243500486
2.223 0.359293531917688
2.425 0.375273593208161
2.626 0.390520312135952
2.827 0.405185268640642
3.02800000000001 0.419351120864291
3.23 0.43313716223246
3.431 0.446446312451957
3.63200000000001 0.4593707275422
3.834 0.472003390615766
4.036 0.484304407252596
4.236 0.488002688313511
4.438 0.48955235548748
4.64 0.491099657792218
4.84099999999999 0.492639030301964
5.042 0.494173038232731
5.244 0.495711826680106
5.44499999999999 0.497235607309031
5.64700000000001 0.498762376164573
5.848 0.500277875600415
6.05 0.501796076490384
6.25 0.503288483119329
6.452 0.504785649440908
6.654 0.506281894511362
6.855 0.507764788877224
7.057 0.509247251040505
7.258 0.510723048573981
7.459 0.512194110244229
7.661 0.513668463750197
7.86199999999999 0.515117548111632
8.06399999999999 0.516568244113482
8.265 0.518004592267568
8.467 0.519440977453482
8.66800000000001 0.5208703061207
8.87 0.522300093394239
9.071 0.523728330321413
9.27200000000001 0.523785539898077
9.473 0.523785539898077
9.676 0.523785539898077
9.877 0.523785539898077
10.078 0.523785539898077
10.279 0.523785539898077
10.48 0.523785539898077
10.682 0.523785539898077
10.884 0.523785539898077
11.085 0.523785539898077
11.287 0.523785539898077
11.488 0.523785539898077
11.689 0.523785539898077
11.89 0.523785539898077
12.092 0.523785539898077
12.293 0.523785539898077
12.495 0.523785539898077
12.696 0.523785539898077
12.898 0.523785539898077
13.099 0.523785539898077
13.3 0.523785539898077
13.502 0.523785539898077
13.703 0.523785539898077
13.905 0.523785539898077
14.106 0.524596154968512
14.308 0.53110035626495
14.509 0.537490125125287
14.71 0.543831247930558
14.912 0.550129220484985
15.113 0.556319279353718
15.315 0.562481941720334
15.517 0.568563377448427
15.718 0.574535827673417
15.919 0.580427248395744
16.121 0.586279642689352
16.321 0.592003816769281
16.523 0.597724591250516
16.724 0.603348533629462
16.926 0.608967688999929
17.127 0.614516925057533
17.329 0.620039464198119
17.53 0.625472503041413
17.732 0.630877086076645
17.933 0.63620899518093
18.134 0.641478102378886
18.336 0.646713522573158
18.538 0.651909486724756
18.738 0.657002801767845
18.94 0.662109175879822
19.141 0.666067201327294
19.343 0.667747720655653
19.544 0.669407373715613
19.746 0.671063012940958
19.947 0.672706982757244
20.148 0.674346862473256
20.349 0.67598070786808
20.552 0.677626235719854
20.753 0.679243032942568
20.955 0.680868500454453
21.155 0.682471252522935
21.357 0.684086968307144
21.558 0.685694928067298
21.76 0.687301569648926
21.962 0.688901074670325
22.163 0.690491302029413
22.364 0.692071846369329
22.565 0.693652111626837
22.767 0.695241472471495
22.969 0.696822580902007
23.17 0.698389778995661
23.371 0.699953498627034
23.573 0.701511900085211
23.775 0.70307044022104
23.975 0.704608088772319
24.177 0.712185255479913
24.378 0.727867798102229
24.58 0.74328132526985
24.781 0.758270463222213
24.983 0.773006981976338
25.184 0.787329840878169
25.384 0.80127028816935
25.587 0.815125905648762
25.788 0.828563379681471
25.989 0.841729687645144
26.191 0.854704423312948
26.392 0.867374841423548
26.593 0.879849084566023
26.795 0.892159673502086
26.997 0.904246091064182
27.198 0.916067135078723
27.398 0.927633214988001
27.6 0.939102997769999
27.801 0.950323314950726
28.003 0.961423810276462
28.204 0.972295910204183
28.406 0.98305273460588
28.607 0.993587245803331
28.81 1.00404620122885
29.01 1.01411847652385
29.213 1.0174398029102
29.413 1.01729868462312
29.614 1.0171797521222
29.816 1.01704909368805
30.017 1.01692212238448
30.219 1.01679151670632
30.419 1.01666985116179
30.621 1.01655449205087
30.823 1.01642721700685
31.024 1.01630558492524
31.225 1.01618484312074
31.427 1.01606235477678
31.629 1.01594643108953
31.87 1.01583290720356
32.072 1.01573575299815
32.273 1.01564410065413
32.475 1.01555387190746
32.676 1.01545481414947
32.877 1.01536452176557
33.079 1.01525822126694
33.28 1.01515622310135
33.482 1.01504708918751
33.682 1.01494204833266
33.885 1.01484540644439
34.087 1.01472329733934
34.287 1.01390275145213
34.489 1.01308552691251
34.691 1.01228262187993
34.892 1.011496755267
35.093 1.01072033755983
35.294 1.00995830167345
35.496 1.00920220017254
35.697 1.00845954847392
35.899 1.0077262592572
36.1 1.00700684818799
36.301 1.00629766341421
36.503 1.00559685976198
36.704 1.00491070437043
36.906 1.00423333164628
37.107 1.00356963822166
37.309 1.00291018508976
37.51 1.00226398585873
37.711 1.00162771682807
37.913 1.00099885359875
38.114 1.00038268070914
38.316 0.999772196921029
38.517 0.999169617601337
38.719 0.998565157292198
38.919 0.997969808013929
39.121 0.998572456282217
39.322 1.00391959466141
39.525 1.00920625030014
39.726 1.01433071405501
39.927 1.01934212124241
40.129 1.02426782501655
40.329 1.02904597652896
40.531 1.03377147505327
40.732 1.03837752630143
40.933 1.04288907700151
41.135 1.04733563120194
41.336 1.05166809537208
41.538 1.05593371315784
41.739 1.06009223478235
41.941 1.06418768280424
42.142 1.06817318212164
42.344 1.07209430335614
42.545 1.07591264299044
42.747 1.07966997230252
42.948 1.08333695571763
43.149 1.08693574744459
43.35 1.09046044193327
43.553 1.09395490603216
43.754 1.0973498838089
43.956 1.10069346446782
44.157 1.10322703956949
44.358 1.10479651757188
44.56 1.10633984865438
44.761 1.10785132755611
44.963 1.10933808423301
45.164 1.1107897821693
45.366 1.11222046187222
45.567 1.11361184473203
45.768 1.11498438998441
45.97 1.11634508169838
46.171 1.11767082176608
46.372 1.11897165063196
46.574 1.12025444210535
46.775 1.12149998980128
46.977 1.12272772103397
47.178 1.12392809147917
47.379 1.12509942866211
47.581 1.12625722461028
47.783 1.12739468018074
47.983 1.12850492775808
48.186 1.12961421191918
48.386 1.13068752472458
48.588 1.13174883263692
48.789 1.13278471447586
48.99 1.13379949381629
49.193 1.13507868827652
49.393 1.13651207928637
49.595 1.13793693477388
49.796 1.13932872218555
49.998 1.14069779995815
50.199 1.14203244772121
50.401 1.14333929271808
50.601 1.14460747270663
50.803 1.14586252772772
51.004 1.14709113167905
51.206 1.14830639675365
51.407 1.14949403315001
51.609 1.15067541053949
51.81 1.15182494176888
52.012 1.15295800356006
52.212 1.154056207339
52.415 1.15515350591057
52.616 1.15620953922225
52.816 1.15723831898682
53.019 1.15826632431247
53.22 1.15927375460827
53.422 1.16026121009956
53.622 1.16121979059126
53.824 1.16217586481495
54.026 1.1631130011599
54.227 1.16364461604514
54.429 1.16403108480089
54.63 1.16440479127101
54.832 1.16477976201816
55.033 1.16514767997561
55.234 1.16550661026849
55.436 1.16586518441166
55.637 1.16621812693713
55.839 1.16655892784527
56.04 1.16689424295555
56.242 1.16722638779915
56.442 1.16754581553026
56.645 1.16786557981574
56.845 1.16817338249037
57.047 1.16848166104286
57.249 1.16879067814767
57.449 1.169085551534
57.651 1.16936824327432
57.852 1.16964243604246
58.055 1.16991611209982
58.255 1.17017978175874
58.457 1.17044332236944
58.658 1.17070331276421
58.86 1.17095791354741
59.061 1.17120818171445
59.263 1.16826811637335
59.463 1.16510797937677
59.665 1.16196413129091
59.866 1.15888068193914
60.068 1.15583008558968
60.27 1.15282553966969
60.471 1.14987887706065
60.672 1.14697879643488
60.874 1.1441099327886
61.075 1.14129770989991
61.276 1.13852870174194
61.479 1.13577432523364
61.679 1.13309737972925
61.881 1.13043665274609
62.083 1.12781544242585
62.284 1.12524742460152
62.485 1.12271930942729
62.687 1.12021526507815
62.888 1.11776349827589
63.089 1.11535103096853
63.291 1.11296150487434
63.492 1.11061944552575
63.694 1.10830122962158
63.894 1.10603868247664
64.096 1.10396699281747
64.298 1.10390398779554
64.499 1.10384688044247
64.701 1.10379245495578
64.902 1.1037375388666
65.103 1.10367671740913
65.305 1.10362274833467
65.506 1.10356822156088
65.708 1.10351237015485
65.909 1.10345997557341
66.11 1.10341274459498
66.312 1.1033606269194
66.514 1.10331159442072
66.715 1.1032687876689
66.916 1.10322299767572
67.118 1.10318041011277
67.319 1.10313558173937
67.521 1.10309512765574
67.722 1.10305587708548
67.924 1.10301836484336
68.125 1.10297807777068
68.326 1.10293617997637
68.527 1.10289356215116
68.728 1.10285301421032
68.931 1.10281348470139
69.132 1.10375040851782
69.334 1.10745285323047
69.534 1.11104792295104
69.735 1.1146248770428
69.936 1.11813031478828
70.139 1.12159515463672
70.34 1.12496367196539
70.542 1.12827723434919
70.742 1.13148882272375
70.944 1.13466864714009
71.146 1.13778106316723
71.346 1.14079617925958
71.549 1.14380412046453
71.749 1.14670987116099
71.951 1.14957895669968
72.152 1.15238666649137
72.354 1.15514304948185
72.556 1.15784346718711
72.757 1.16047130431092
72.958 1.16304782077241
73.159 1.16557475836719
73.361 1.16807108465612
73.562 1.17050207309866
73.764 1.1729028267072
73.965 1.17524088347223
74.167 1.17527978227133
74.368 1.17242964866445
74.569 1.16962390353759
74.772 1.16683681806803
74.973 1.16411898319845
75.174 1.16144479831029
75.375 1.15880742479134
75.576 1.15621182429689
75.778 1.15364055163885
75.979 1.15111687501895
76.181 1.1486193819204
76.382 1.14617634078007
76.584 1.14376040552031
76.785 1.14140048939448
76.986 1.13907833007774
77.188 1.13677980878799
77.389 1.1345251850526
77.591 1.13229596180756
77.792 1.13010691362903
77.993 1.12795353652717
78.195 1.12582059675482
78.396 1.12373321734435
78.597 1.12167982689171
78.799 1.11964433159688
79 1.11765185481545
79.202 1.11946147764753
79.403 1.12384339561852
79.603 1.12810396989759
79.805 1.13232690195885
80.006 1.13644377140386
80.209 1.14053119918345
80.41 1.14450237929372
80.611 1.1483935833358
80.813 1.15221770649386
81.014 1.1559397071213
81.216 1.15960192386193
81.416 1.16314685161589
81.619 1.16667458256236
81.819 1.17007659271293
82.021 1.17343651977021
82.223 1.17672103547264
82.423 1.17991091877003
82.625 1.18307414953632
82.826 1.18616120261121
83.028 1.18920148075431
83.229 1.19216062181644
83.431 1.195076375894
83.632 1.19791365130219
83.833 1.20069100486679
84.067 1.20385246204643
84.269 1.20402232590796
84.471 1.2040264736809
84.672 1.20402273955023
84.873 1.20402850522961
85.074 1.20403489101128
85.276 1.20404221497182
85.477 1.20404675624563
85.679 1.20404217698284
85.88 1.20404285403364
86.082 1.20405133716887
86.283 1.20405300186128
86.483 1.20405926929568
86.686 1.20406654086431
86.887 1.20407323732098
87.089 1.20407594481235
87.289 1.20408220485281
87.491 1.20409272973276
87.693 1.20410229537548
87.894 1.20410590159448
88.095 1.20411115950296
88.297 1.20411900314872
88.498 1.20412787305966
88.7 1.20413685677619
88.901 1.20414608567773
89.103 1.20364080205156
89.304 1.19956157009553
89.506 1.19552256503581
89.707 1.19156303603938
89.908 1.18766174093065
90.11 1.18379893108948
90.311 1.18001197505109
90.512 1.17628088629301
90.714 1.17258664891789
90.916 1.16894701160581
91.117 1.16537890283666
91.319 1.16184595073698
91.519 1.15839942792843
91.721 1.15496955187876
91.923 1.15159051380225
92.123 1.14829411718466
92.325 1.14501377691368
92.526 1.14179787491368
92.728 1.13861369380605
92.929 1.13549216639119
93.13 1.13241666235631
93.332 1.12937143637298
93.533 1.12638597256054
93.735 1.12343000586374
93.936 1.12053197165902
94.137 1.11779673969071
94.339 1.11536763457058
94.54 1.11298709980374
94.742 1.11063036871753
94.943 1.10832236936663
95.144 1.10605343005469
95.346 1.10380728913735
95.548 1.10159450910045
95.748 1.09943143989955
95.951 1.09726955868907
96.152 1.09516194425252
96.353 1.09308634415923
96.555 1.09103389616575
96.755 1.08903193168642
96.958 1.08703168077653
97.158 1.08509234147511
97.361 1.08315044251429
97.561 1.08126779395675
97.764 1.07938583856988
97.965 1.07754909625734
98.166 1.07574358389387
98.368 1.07395291886313
98.569 1.07219412954666
98.77 1.07046525528584
98.972 1.06875156463544
99.173 1.06726337066036
99.375 1.06601712404994
99.576 1.06479120618463
99.777 1.06358602950934
99.978 1.06239842143412
100.181 1.06121050421209
100.381 1.06006000017785
100.584 1.05890934526821
100.785 1.0577945890649
100.986 1.0567026012498
101.188 1.05561971399094
101.389 1.05455818963347
101.591 1.05351066633923
101.791 1.05248989721425
101.993 1.05146700534876
102.194 1.05045847493038
102.396 1.0494584323257
102.596 1.04848300686103
102.799 1.04751320430084
102.999 1.04656877334218
103.202 1.04563222191087
103.403 1.04472238764033
103.604 1.04382278199075
103.806 1.04293610974404
104.007 1.0420663933236
104.208 1.04065442374881
104.41 1.03896311173371
104.611 1.03730496151331
104.813 1.03566319804202
105.014 1.03405372129453
105.215 1.03246789453724
105.417 1.03089768426458
105.618 1.02935828349424
105.82 1.02783404120822
106.021 1.02633976472207
106.222 1.02486746097164
106.424 1.02340970184623
106.626 1.02197353082619
106.827 1.02056552843213
107.029 1.0191714621998
107.229 1.01781148784967
107.432 1.01645150063581
107.633 1.01512489495084
107.835 1.01381131504511
108.036 1.01252348982328
108.237 1.01125467432881
108.438 1.01000450946794
108.64 1.00876667422567
108.841 1.00755318539785
109.042 1.00635756327162
109.244 1.01080710487567
109.445 1.01639617955456
109.647 1.02190714296801
109.848 1.02727059557853
110.05 1.03255463871157
110.251 1.0377020788081
110.453 1.04276776016832
110.654 1.04770006140113
110.855 1.05252774780227
111.056 1.05724987558896
111.258 1.06189769542917
111.459 1.06642290712625
111.661 1.07087925407133
111.862 1.07522233587451
112.063 1.07947767553358
112.265 1.08366386379578
112.467 1.08775991658751
112.668 1.09175421911309
112.869 1.09566012897825
113.071 1.09951250902027
113.272 1.10326606759315
113.474 1.10696456510672
113.675 1.11056847693611
113.876 1.11410349907084
114.078 1.11758301868822
114.28 1.11784373541332
114.481 1.11811473055459
114.682 1.11838151029173
114.884 1.11864178989833
115.085 1.11889918605789
115.286 1.11914891946386
115.487 1.11939986669811
115.69 1.11964562359236
115.89 1.11988393336835
116.092 1.12012354177523
116.294 1.12035932334024
116.496 1.12058765773181
116.697 1.12081375075629
116.898 1.12102752051005
117.1 1.1212358485796
117.301 1.12144088132172
117.502 1.12164435349089
117.703 1.12184088750568
117.905 1.12203824958108
118.107 1.12222778636433
118.307 1.12241078894811
118.509 1.12259745103331
118.71 1.1227737020402
118.912 1.1229503748942
119.114 1.12261110686475
119.315 1.11972331696608
119.516 1.11687809871489
119.718 1.1140609215904
119.919 1.11129904196236
120.121 1.10856438155871
120.322 1.10588340951229
120.523 1.10324196305611
120.725 1.1006265737789
120.926 1.09806251651879
121.127 1.09553628362485
121.328 1.09304722498843
121.53 1.0905827788706
121.731 1.08816667631145
121.933 1.08577432890935
122.134 1.08342898809003
122.335 1.08111820188489
122.537 1.07883021528357
122.738 1.0765871928173
122.941 1.07435530240285
123.142 1.07217814379223
123.344 1.07002243374916
123.545 1.0679090535598
123.747 1.06581653334734
123.947 1.06377531779269
124.148 1.06243720280059
124.35 1.06243702180258
124.551 1.06243700489709
124.753 1.06243593577465
124.954 1.06243404492691
125.155 1.06243328146672
125.358 1.06243018552594
125.558 1.06242911375532
125.76 1.06242730863144
125.962 1.06242511748994
126.163 1.06242511187635
126.365 1.06242502507788
126.566 1.06242461861889
126.767 1.06242461861889
126.968 1.06242390181507
127.17 1.06242341410237
127.372 1.0624230320145
127.573 1.06242206639352
127.775 1.06242158534322
127.976 1.06242058090364
128.176 1.06241868777226
128.379 1.0624182170202
128.58 1.06241805711607
128.782 1.06241805283623
128.982 1.06241805283623
129.184 1.06350685733345
129.384 1.0655381558214
129.586 1.06756230033203
129.788 1.06954837238894
129.99 1.07149959258373
130.19 1.07339654492928
130.393 1.07528608773961
130.593 1.07710112605676
130.795 1.07889129593072
130.997 1.08065398997384
131.198 1.08236637211792
131.398 1.08404167824103
131.601 1.08571690582908
131.801 1.08734110555965
132.004 1.08896184216556
132.205 1.09054093827874
132.406 1.09209251808922
132.607 1.09361689044319
132.809 1.0951148006945
133.01 1.09656559126278
133.211 1.09798727860892
133.414 1.09939850342424
133.615 1.10076726749894
133.816 1.10211694641509
134.017 1.10344268471828
134.219 1.13257894115945
134.42 1.17381548723613
134.622 1.21317367209025
134.823 1.25042072446785
135.024 1.2861335423963
135.226 1.32081706727831
135.427 1.35418273521436
135.628 1.38648048494596
135.829 1.41781540330221
136.031 1.4483926954037
136.233 1.47808236086993
136.433 1.50664954436624
136.636 1.53495711898336
136.837 1.56236255350334
137.038 1.58916901451845
137.239 1.61541426783753
137.442 1.64139847447027
137.642 1.66648201535854
137.844 1.69133326682407
138.045 1.71558874365988
138.247 1.73952524851045
138.448 1.74058158505902
138.649 1.7405803443452
138.851 1.7405803920571
139.051 1.74057987759106
139.253 1.73602412220761
139.455 1.7314348653006
139.656 1.7268786866333
139.857 1.72232709914959
140.1 1.71684359085833
140.301 1.71231388844174
140.503 1.7077685049215
140.704 1.7032537645711
140.905 1.69875054977739
141.107 1.69422864171607
141.307 1.68976351246607
141.51 1.68524011285562
141.71 1.68079781193881
141.913 1.67629431694965
142.113 1.67186041084256
142.315 1.66738526351977
142.516 1.6629382768949
142.718 1.65847408956356
142.919 1.65404532446616
143.12 1.6496246939334
143.322 1.6451908502625
143.522 1.64081624407579
143.725 1.63637859281476
143.926 1.63199651398711
144.127 1.63298077550204
144.329 1.65440049415063
144.529 1.67522521229447
144.731 1.69589999331316
144.933 1.71624587435738
145.134 1.73617747845668
145.335 1.74048559947344
145.537 1.74048579214698
145.738 1.74048511527365
145.94 1.74048410965755
146.141 1.74048522384195
146.342 1.74048527438456
146.544 1.74048518185016
146.745 1.74048500955056
146.947 1.7404856653759
147.148 1.74048575508637
147.349 1.74048497202125
147.552 1.74048668101918
147.752 1.74048671556237
147.954 1.74048711144127
148.155 1.74048390803091
148.357 1.74048498202759
148.557 1.74048488830048
148.759 1.74048342223363
148.96 1.74048384784526
149.162 1.74032491873911
149.363 1.7403262905773
149.565 1.7403267630097
149.766 1.74032632901021
149.968 1.74032657338894
150.169 1.74032744818702
150.371 1.74032761428178
150.572 1.74032761037481
150.774 1.7403275072979
150.975 1.74032664869317
151.175 1.74032676253512
151.377 1.74032547338466
151.578 1.74032635396617
151.78 1.7403252126192
151.981 1.74032634112221
152.183 1.74032608930614
152.384 1.74032775197214
152.586 1.74032738138165
152.787 1.74032703375324
152.989 1.74032841314456
153.19 1.74032789008775
153.391 1.74032781217215
153.593 1.74032802431173
153.793 1.74032799348836
153.995 1.74032630253742
154.197 1.73826150375018
154.399 1.73505866887761
154.6 1.73187882412354
154.802 1.7286993079533
155.002 1.72556444046434
155.204 1.72240490197387
155.405 1.71926431218151
155.606 1.71613445833722
155.809 1.71298116370337
156.009 1.70987637373429
156.211 1.70675257254939
156.413 1.70363606386968
156.614 1.70054190591107
156.815 1.69745885161875
157.017 1.6943723726601
157.219 1.69130052528394
157.419 1.6882686322821
157.621 1.68521451721334
157.823 1.68216880410965
158.024 1.67914487703389
158.225 1.67611672222808
158.426 1.67309620853308
158.628 1.67006727318904
158.829 1.66706169836391
159.031 1.66405253036942
159.231 1.65967326507942
159.433 1.65482441500093
159.635 1.64998301833847
159.837 1.64514959237796
160.037 1.6403691064764
160.239 1.63554808054321
160.44 1.6307584214021
160.642 1.62594915920588
160.843 1.62117073446074
161.045 1.61637717684048
161.246 1.61161208807224
161.448 1.60683385001223
161.649 1.60208946799365
161.85 1.59735311983206
162.052 1.59260556624274
162.253 1.58788681298797
162.454 1.58317423334521
162.656 1.57844742406996
162.858 1.57372485077119
163.058 1.56905029243146
163.26 1.56434454825745
163.462 1.55964375567772
163.663 1.55497355424394
163.864 1.55031768612473
164.066 1.54564601791742
164.266 1.53997000692915
164.468 1.53416537496741
164.669 1.5283952180642
164.871 1.52260160617213
165.073 1.51681410477159
165.274 1.51106093225664
165.475 1.50531290487278
165.677 1.4995420768011
165.878 1.49381500306844
166.08 1.4880821560167
166.281 1.48239953119954
166.483 1.47671033110252
166.684 1.47107222563032
166.886 1.46542851220836
167.086 1.45986309382558
167.287 1.45429208759965
167.49 1.44868860512849
167.691 1.44316260611365
167.893 1.4376315731942
168.093 1.43217680096371
168.296 1.42666203566777
168.496 1.42125008559457
168.698 1.41580489999409
168.899 1.41040800695151
169.101 1.40504468158423
169.303 1.40002524030587
169.504 1.39505152757188
169.704 1.39012341533174
169.906 1.38516753015162
170.108 1.38023450048568
170.31 1.37532178039853
170.511 1.37045361236027
170.712 1.36560739878069
170.914 1.36075579954472
171.115 1.35594757538812
171.317 1.35113599653761
171.518 1.34636840454176
171.719 1.34161931797844
171.92 1.33689351635075
172.122 1.33216708002652
172.324 1.32746013341673
172.525 1.3227967974741
172.727 1.31812914405088
172.927 1.31352941250998
173.129 1.30890075599426
173.331 1.3042901356481
173.532 1.29972123245883
173.733 1.29517371593983
173.935 1.29062161195238
174.136 1.28604742363193
174.337 1.28132794754525
174.539 1.27660286726626
174.74 1.27192162104223
174.942 1.26723634262488
175.143 1.26259244268912
175.345 1.25794626024485
175.546 1.25334396032704
175.748 1.24873745138321
175.949 1.24420938175282
176.151 1.23972841030818
176.351 1.2353554997254
176.553 1.23100406927585
176.755 1.2267181583122
176.955 1.22253541232716
177.157 1.21837342445875
177.358 1.21429456453339
177.56 1.21025595545882
177.761 1.20630117780353
177.963 1.20238551473176
178.164 1.19854669795986
178.366 1.19474526074027
178.567 1.19101961959312
178.768 1.18735043051698
178.97 1.18371825575671
179.171 1.18378910647021
179.372 1.18923330392297
179.574 1.19458420297804
179.775 1.19977498742024
179.977 1.20488862427998
180.178 1.20988677280269
180.379 1.21476827512837
180.581 1.21956567473478
180.782 1.22423180051882
180.984 1.22881611536732
181.185 1.23327665982517
181.387 1.23766218493505
181.588 1.24192641390788
181.79 1.24611887760253
181.991 1.25018261395959
182.193 1.25419828317606
182.394 1.25818477277918
182.594 1.2621167089489
182.797 1.26605974553852
182.997 1.26989788490049
183.199 1.27373507480299
183.4 1.27751438836692
183.602 1.28125388395748
183.803 1.28493269528772
184.005 1.28859096458611
184.206 1.29498207256109
184.408 1.303751947562
184.609 1.31236123840734
184.811 1.32096001126148
185.012 1.32945152896401
185.214 1.33785666088906
185.415 1.34611275881475
185.617 1.35429204484479
185.818 1.36231930242984
186.02 1.37026180568993
186.22 1.37802817403908
186.422 1.38577460290871
186.623 1.39337711152898
186.825 1.40092466038644
187.026 1.40834193660844
187.228 1.41569892115328
187.429 1.42293598956138
187.631 1.43011913656903
187.832 1.43716784594058
188.034 1.44415795320967
188.235 1.45102659579533
188.436 1.45781845668612
188.638 1.46454599061525
188.839 1.47115815812565
189.041 1.47771744889182
189.242 1.47735136847861
189.444 1.47535854050576
189.645 1.47338516618646
189.846 1.47142215789676
190.048 1.46946465768026
190.249 1.46753074492955
190.451 1.46559403248474
190.652 1.46368111345471
190.854 1.46176769549688
191.055 1.45987610833884
191.297 1.45762056730171
191.498 1.45575475069406
191.7 1.45389557243966
191.901 1.45205499637808
192.102 1.45022533375311
192.303 1.44841358452238
192.505 1.44660520960345
192.707 1.44480578108187
192.908 1.4430306788593
193.11 1.44125053482504
193.311 1.43949079000335
193.512 1.43774430403105
193.714 1.43600675848189
193.915 1.43428981291935
194.117 1.43198034973448
194.318 1.42679390921544
194.52 1.42160388397997
194.721 1.41646213346276
194.922 1.41134051436091
195.123 1.40623961756892
195.324 1.40116027580462
195.527 1.39605267405167
195.727 1.39104135180877
195.929 1.38600051642394
196.131 1.38098135840314
196.333 1.37598410199998
196.533 1.37105500098118
196.735 1.36609914320008
196.936 1.3611882871221
197.137 1.35629862343082
197.34 1.35137910655062
197.541 1.346528255301
197.743 1.34167462183555
197.943 1.33688888819112
198.146 1.33205116077353
198.346 1.32730597167175
198.549 1.32251142584244
198.749 1.31780754920176
198.951 1.31307732936234
199.152 1.30903723242401
199.353 1.30614634037147
199.555 1.30325428943562
199.756 1.3003927009354
199.958 1.29753428473252
};
\addlegendentry{\small Buffer capacitor voltage}
\addplot [semithick, red]
table {%
-9.99475 1.74
209.95575 1.74
};
\addlegendentry{\small Target voltage (MCU)}
\end{axis}

\end{tikzpicture}

%% file: tikz/multi_tone_m1_80_energy_buffer.tex
\begin{tikzpicture}

\definecolor{darkgray176}{RGB}{176,176,176}
\definecolor{darkorange25512714}{RGB}{255,127,14}
\definecolor{lightgray204}{RGB}{204,204,204}
\definecolor{steelblue31119180}{RGB}{31,119,180}

\begin{axis}[
width=\columnwidth,
height=0.55\columnwidth,
legend cell align={left},
line cap = round,
legend style={
  fill opacity=0.8,
  draw opacity=1,
  text opacity=1,
  at={(0.97,0.03)},
  anchor=south east,
  draw=lightgray204
},
tick align=outside,
tick pos=left,
x grid style={darkgray176},
xlabel={Time [s]},
xmajorgrids,
xmin=-4.995, xmax=104.939,
xtick style={color=black},
y grid style={darkgray176},
ylabel={Voltage [V]},
ymajorgrids,
ymin=-0.090661053757336, ymax=2.1,
ytick style={color=black}
]
\addplot [semithick, steelblue31119180]
table {%
0.00199999999999534 1.763
0.102999999999994 1.548
0.204000000000008 1.13
0.305000000000007 1.254
0.405000000000001 1.014
0.506 0.733
0.606999999999999 1.421
0.707000000000008 1.474
0.808000000000007 1.551
0.909000000000006 1.994
1.01000000000001 0.885
1.11 0.796
1.211 0.793
1.312 1.488
1.413 1.724
1.51300000000001 1.731
1.613 1.65
1.715 1.521
1.815 1.335
1.916 1.308
2.01600000000001 1.451
2.117 1.771
2.21900000000001 1.524
2.318 1.48
2.419 1.454
2.52 1.118
2.621 1.167
2.721 1.06
2.822 1.794
2.923 1.857
3.024 2.032
3.124 1.835
3.224 1.638
3.32600000000001 0.976
3.426 1.158
3.527 1.612
3.628 1.792
3.72799999999999 1.034
3.82900000000001 0.978
3.93000000000001 1.068
4.03100000000001 1.192
4.131 1.56
4.232 1.686
4.333 1.982
4.43300000000001 1.649
4.53400000000001 1.837
4.63500000000001 0.851
4.736 0.852
4.837 1.572
4.937 1.552
5.038 1.655
5.139 1.563
5.239 1.561
5.34 1.607
5.441 1.151
5.54000000000001 1.108
5.642 1.23
5.742 1.582
5.84400000000001 1.541
5.944 0.951
6.045 1.005
6.146 0.878
6.246 0.864
6.346 1.06
6.44800000000001 1.612
6.54900000000001 2.035
6.648 2.044
6.75 1.766
6.851 0.842
6.95100000000001 1.047
7.05200000000001 1.517
7.15300000000001 1.543
7.253 1.71
7.354 0.713
7.455 1.316
7.55500000000001 1.422
7.657 1.345
7.756 1.072
7.857 1.523
7.958 1.499
8.06 2.085
8.16 1.05
8.26000000000001 0.906
8.361 1.479
8.461 1.445
8.562 1.422
8.663 1.604
8.764 1.577
8.864 1.725
8.96600000000001 0.942
9.066 1.136
9.167 1.218
9.267 1.224
9.369 1.51
9.46900000000001 1.026
9.57000000000001 1.201
9.67 0.959
9.771 0.967
9.871 1.088
9.973 1.569
10.073 2.021
10.174 2.076
10.274 1.914
10.376 0.869
10.475 0.855
10.577 1.56
10.678 1.534
10.778 1.098
10.879 0.94
10.979 0.922
11.081 1.335
11.18 1.333
11.281 1.318
11.382 1.415
11.484 1.984
11.583 1.974
11.685 0.945
11.785 0.925
11.886 1.26
11.987 1.292
12.088 1.665
12.189 1.647
12.289 1.568
12.389 1.656
12.49 0.801
12.591 1.564
12.691 1.317
12.792 1.295
12.893 1.787
12.993 1.661
13.094 1.161
13.195 1.107
13.296 1.312
13.397 1.71
13.498 1.633
13.598 2.001
13.698 1.916
13.8 1.062
13.9 0.855
14.001 0.703
14.101 0.797
14.202 1.575
14.303 1.218
14.403 1.067
14.504 1.019
14.605 1.413
14.706 1.297
14.807 1.037
14.907 1.209
15.008 1.965
15.108 1.968
15.209 2.043
15.309 1.177
15.411 0.996
15.511 1.166
15.612 1.404
15.713 1.854
15.813 1.789
15.914 1.478
16.015 1.611
16.115 1.045
16.216 1.23
16.317 1.597
16.417 1.767
16.518 1.462
16.62 1.187
16.72 0.985
16.82 1.194
16.921 1.134
17.022 1.447
17.123 1.375
17.223 2.002
17.324 2.047
17.425 1.031
17.525 0.735
17.626 0.551
17.727 1.699
17.827 1.711
17.927 1.497
18.029 1.378
18.13 1.333
18.23 1.34
18.331 1.452
18.432 1.414
18.532 1.383
18.633 2.006
18.734 1.686
18.835 1.628
18.935 1.337
19.036 1.239
19.137 1.402
19.237 1.862
19.337 1.772
19.439 1.416
19.54 1.578
19.64 1.74
19.741 0.811
19.841 0.828
19.942 1.515
20.043 1.645
20.143 1.656
20.244 1.439
20.345 1.019
20.447 0.954
20.546 0.978
20.646 1.444
20.748 1.863
20.849 1.905
20.95 1.981
21.049 2.018
21.151 1.006
21.252 1.598
21.352 1.436
21.453 1.309
21.553 1.309
21.654 1.205
21.755 1.208
21.856 1.357
21.956 1.248
22.058 2.002
22.157 2.078
22.259 2.098
22.359 1.708
22.46 1.489
22.561 1.147
22.662 1.157
22.762 1.746
22.862 1.782
22.964 1.353
23.064 1.464
23.165 1.482
23.266 0.878
23.367 0.843
23.466 0.939
23.567 1.555
23.668 1.788
23.77 1.494
23.869 1.112
23.97 1.037
24.071 1.114
24.172 1.194
24.272 1.76
24.373 1.629
24.474 1.879
24.574 1.283
24.675 1.117
24.776 1.208
24.877 1.492
24.977 1.732
25.078 1.381
25.179 1.382
25.28 1.098
25.38 1.298
25.481 1.078
25.582 1.116
25.682 1.812
25.783 1.887
25.883 2.057
25.984 1.593
26.086 0.946
26.187 0.995
26.287 1.585
26.388 1.581
26.488 1.69
26.588 1.323
26.689 1.334
26.79 1.258
26.891 0.998
26.991 1.687
27.092 1.621
27.193 1.626
27.294 1.882
27.393 1.662
27.495 1.091
27.596 1.077
27.697 1.458
27.796 1.347
27.898 1.743
27.998 1.701
28.099 1.517
28.2 1.298
28.301 1.225
28.401 0.858
28.502 1.465
28.603 1.432
28.704 1.351
28.805 1.182
28.905 1.35
29.005 1.387
29.106 1.207
29.207 1.54
29.308 1.807
29.409 1.879
29.509 1.795
29.61 0.76
29.71 0.843
29.811 0.98
29.912 1.494
30.012 1.715
30.113 0.993
30.214 1.052
30.314 1.181
30.415 1.331
30.516 1.119
30.616 1.661
30.717 1.528
30.818 1.893
30.919 1.819
31.02 1.096
31.12 1.092
31.221 1.35
31.322 1.248
31.422 1.797
31.524 1.851
31.624 1.827
31.724 1.474
31.826 1.066
31.926 0.963
32.026 1.345
32.127 1.456
32.229 1.485
32.329 1.026
32.429 1.097
32.53 1.15
32.631 1.29
32.732 1.479
32.832 1.699
32.933 1.892
33.034 1.782
33.134 1.8
33.235 0.731
33.336 1.036
33.436 1.088
33.537 1.468
33.638 0.945
33.739 0.997
33.839 1.151
33.94 1.367
34.041 1.149
34.141 1.554
34.242 1.7
34.343 1.899
34.444 1.897
34.544 1.197
34.645 1.16
34.746 1.237
34.846 1.486
34.947 1.815
35.047 1.891
35.148 1.798
35.25 1.632
35.35 1.599
35.451 1.296
35.551 1.165
35.652 1.606
35.753 1.703
35.853 1.583
35.955 0.92
36.055 0.834
36.155 0.864
36.257 1.689
36.357 1.646
36.458 1.603
36.558 2.023
36.659 1.946
36.76 1.746
36.86 0.657
36.961 0.854
37.062 1.669
37.163 1.124
37.262 1.527
37.364 1.174
37.465 1.238
37.566 1.271
37.666 1.617
37.766 1.7
37.868 1.978
37.968 1.831
38.069 1.673
38.169 1.392
38.27 1.271
38.371 1.203
38.471 1.553
38.572 1.85
38.673 1.942
38.774 1.698
38.874 1.782
38.974 1.668
39.075 1.429
39.176 1.188
39.277 1.39
39.377 1.687
39.479 0.851
39.579 0.878
39.679 1.135
39.781 1.164
39.881 1.034
39.983 1.584
40.083 1.55
40.184 1.991
40.284 2.004
40.385 0.913
40.486 0.699
40.587 0.718
40.687 1.393
40.787 1.428
40.888 1.072
40.99 1.314
41.089 1.152
41.19 1.273
41.292 1.476
41.392 1.563
41.492 1.607
41.593 1.922
41.694 1.923
41.795 1.162
41.895 1.045
41.995 1.086
42.097 1.432
42.197 1.372
42.297 1.823
42.399 1.655
42.5 1.727
42.6 1.02
42.701 1.216
42.802 1.374
42.901 1.409
43.003 1.719
43.103 1.532
43.205 1.166
43.304 1.191
43.406 1.219
43.507 1.566
43.607 1.717
43.708 2.1
43.808 1.697
43.909 1.858
44.01 0.888
44.11 0.738
44.212 1.526
44.312 1.549
44.412 1.339
44.513 0.921
44.614 1.308
44.715 1.341
44.816 1.401
44.916 1.263
45.017 1.95
45.118 1.838
45.218 1.825
45.319 1.975
45.42 1.31
45.52 1.199
45.621 1.102
45.722 1.491
45.823 1.628
45.924 1.738
46.024 1.478
46.124 1.546
46.226 1.118
46.327 1.325
46.427 1.476
46.528 1.452
46.629 1.316
46.73 1.07
46.83 1.039
46.931 1.259
47.032 1.378
47.131 1.594
47.232 1.642
47.334 1.942
47.435 1.196
47.534 1.102
47.635 0.861
47.737 1.541
47.838 1.578
47.937 1.683
48.039 1.191
48.139 1.123
48.239 1.47
48.341 1.354
48.442 1.528
48.542 1.602
48.643 1.987
48.743 2.069
48.844 1.624
48.946 1.719
49.045 1.062
49.146 1.095
49.247 1.481
49.347 1.388
49.448 1.635
49.55 1.405
49.65 1.221
49.75 1.155
49.851 1.048
49.953 1.323
50.052 1.363
50.153 1.607
50.254 1.134
50.356 1.305
50.455 1.326
50.557 1.224
50.658 1.232
50.758 1.889
50.858 1.959
50.959 1.925
51.06 0.935
51.161 1.398
51.261 1.448
51.362 1.609
51.463 1.381
51.563 1.2
51.664 1.341
51.765 1.504
51.866 1.512
51.966 1.616
52.067 1.521
52.168 1.981
52.268 1.565
52.369 1.776
52.469 1.831
52.571 1.09
52.672 1.152
52.771 1.251
52.872 1.556
52.974 1.366
53.074 1.389
53.174 1.381
53.275 1.153
53.376 1.025
53.477 1.708
53.577 1.488
53.678 1.633
53.779 1.537
53.88 1.194
53.98 1.449
54.081 1.416
54.182 1.445
54.282 1.947
54.383 1.832
54.483 1.793
54.585 1.354
54.685 1.176
54.786 1.095
54.887 1.229
54.987 1.371
55.088 1.344
55.189 1.279
55.29 1.199
55.389 1.199
55.49 1.487
55.624 1.804
55.729 1.07
55.829 1.086
55.93 1.433
56.031 1.696
56.131 1.642
56.232 1.647
56.333 1.909
56.434 1.754
56.534 1.799
56.635 1.727
56.736 1.105
56.837 1.227
56.937 1.417
57.038 1.516
57.138 1.776
57.238 1.369
57.34 1.128
57.441 1.079
57.542 1.073
57.642 1.465
57.742 1.578
57.843 1.194
57.944 0.99
58.045 1.295
58.145 1.424
58.247 1.357
58.347 1.443
58.447 1.579
58.548 1.847
58.649 1.054
58.75 1.014
58.85 1.085
58.951 1.458
59.052 1.33
59.153 1.275
59.253 1.318
59.354 1.475
59.455 1.502
59.556 1.687
59.656 1.649
59.757 1.886
59.858 1.928
59.958 1.992
60.059 1.863
60.16 1.714
60.261 1.203
60.361 1.371
60.462 1.47
60.562 1.506
60.663 1.502
60.764 1.338
60.864 1.454
60.965 1.2
61.067 1.115
61.166 1.466
61.268 1.393
61.368 1.571
61.468 1.295
61.57 1.541
61.67 1.457
61.771 1.337
61.872 1.439
61.972 1.52
62.073 1.59
62.174 1.532
62.274 1.887
62.375 1.23
62.476 1.29
62.577 1.452
62.678 1.421
62.779 1.328
62.878 1.206
62.98 1.17
63.081 1.504
63.181 1.585
63.282 1.742
63.382 1.835
63.483 1.86
63.584 1.572
63.685 1.587
63.785 1.147
63.886 1.119
63.986 1.191
64.087 1.498
64.188 1.507
64.288 1.581
64.389 1.573
64.489 1.242
64.59 1.266
64.691 1.231
64.792 1.617
64.892 1.454
64.994 1.591
65.094 1.438
65.195 1.383
65.295 1.38
65.397 1.566
65.497 1.386
65.597 1.884
65.698 1.681
65.799 1.692
65.9 1.34
66.001 1.291
66.101 1.406
66.202 1.249
66.302 1.385
66.403 1.319
66.504 1.03
66.605 1.221
66.705 1.336
66.806 1.518
66.907 1.783
67.007 1.884
67.108 1.737
67.209 1.618
67.31 1.725
67.41 0.959
67.511 1.252
67.612 1.437
67.713 1.155
67.813 1.188
67.914 1.407
68.015 1.353
68.115 1.458
68.216 1.353
68.317 1.64
68.417 1.617
68.519 1.487
68.619 1.365
68.719 1.465
68.82 1.142
68.921 1.248
69.022 1.406
69.122 1.41
69.223 1.943
69.324 1.83
69.425 1.508
69.525 1.188
69.626 1.565
69.727 1.516
69.827 1.439
69.927 1.299
70.028 1.252
70.13 1.174
70.23 1.361
70.33 1.515
70.432 1.847
70.532 1.84
70.633 1.836
70.734 1.682
70.834 1.652
70.935 1.021
71.036 1.119
71.136 1.215
71.237 1.216
71.338 1.388
71.439 1.262
71.539 1.351
71.64 1.414
71.74 1.384
71.841 1.392
71.941 1.755
72.043 1.412
72.144 1.554
72.244 1.484
72.345 1.336
72.446 1.428
72.547 1.55
72.647 1.801
72.748 1.877
72.849 1.929
72.949 1.692
73.05 1.74
73.15 1.377
73.251 1.283
73.352 1.33
73.453 1.336
73.554 1.239
73.654 1.272
73.755 1.075
73.855 1.438
73.956 1.551
74.057 1.824
74.158 1.809
74.258 1.92
74.358 1.739
74.46 1.564
74.56 1.047
74.661 1.079
74.762 1.235
74.863 1.374
74.963 0.954
75.064 1.353
75.164 1.304
75.266 1.313
75.365 1.668
75.467 1.814
75.568 1.438
75.668 1.399
75.769 1.653
75.87 1.613
75.97 1.575
76.07 1.491
76.171 1.69
76.272 1.703
76.372 1.876
76.474 1.586
76.574 1.792
76.675 1.302
76.775 1.312
76.877 1.287
76.977 1.359
77.078 1.471
77.179 1.019
77.28 0.892
77.379 0.873
77.481 1.529
77.582 1.744
77.682 1.72
77.783 1.85
77.883 1.785
77.984 1.712
78.086 1.139
78.187 1.404
78.286 1.194
78.387 1.154
78.488 1.34
78.588 1.162
78.69 1.418
78.79 1.26
78.891 1.828
78.991 1.845
79.092 1.591
79.193 1.531
79.293 1.342
79.394 1.77
79.495 1.472
79.596 1.559
79.696 1.652
79.798 1.463
79.897 1.648
79.998 1.857
80.099 1.577
80.2 1.687
80.3 1.305
80.401 1.344
80.503 1.298
80.603 1.412
80.703 0.915
80.804 0.947
80.905 1.048
81.005 1.575
81.106 1.715
81.207 1.795
81.308 1.813
81.408 1.765
81.509 1.184
81.61 1.294
81.711 1.328
81.81 1.463
81.912 1.117
82.012 0.932
82.113 1.288
82.214 1.605
82.314 1.479
82.415 1.803
82.517 1.745
82.617 1.591
82.718 1.506
82.818 1.317
82.919 1.673
83.019 1.458
83.12 1.233
83.221 1.28
83.321 1.434
83.422 1.462
83.523 1.452
83.624 1.441
83.725 1.201
83.825 1.193
83.927 1.632
84.027 1.523
84.127 1.392
84.228 1.215
84.329 1.199
84.429 1.623
84.53 1.723
84.631 1.868
84.732 1.756
84.832 1.766
84.933 1.648
85.034 1.709
85.135 1.398
85.235 1.3
85.335 1.225
85.437 1.277
85.537 1.007
85.638 1.479
85.739 1.193
85.84 1.352
85.94 1.34
86.04 1.749
86.141 1.82
86.243 1.546
86.343 1.479
86.443 1.455
86.545 1.534
86.645 1.287
86.746 1.369
86.847 1.398
86.947 1.62
87.048 1.371
87.149 1.425
87.25 1.377
87.35 1.453
87.451 1.74
87.552 1.563
87.652 1.32
87.753 1.365
87.853 1.398
87.955 1.211
88.055 1.702
88.156 1.587
88.257 1.698
88.358 1.951
88.458 1.674
88.559 1.692
88.66 1.414
88.76 1.261
88.86 1.402
88.961 1.463
89.063 1.306
89.162 1.357
89.264 0.712
89.365 1.035
89.466 1.585
89.566 1.724
89.667 1.686
89.767 1.613
89.868 1.515
89.97 1.432
90.069 1.587
90.17 1.488
90.271 1.327
90.372 1.256
90.472 1.475
90.573 1.538
90.674 1.482
90.774 1.402
90.876 1.627
90.976 1.481
91.077 1.589
91.177 1.462
91.278 1.256
91.379 1.554
91.479 1.41
91.581 1.355
91.681 1.68
91.782 1.654
91.882 1.847
91.983 1.795
92.083 1.77
92.185 1.664
92.286 1.203
92.387 1.325
92.486 1.443
92.587 1.278
92.688 1.283
92.789 1.207
92.89 1.231
92.99 1.703
93.091 1.729
93.192 1.786
93.291 1.714
93.393 1.835
93.493 1.687
93.595 1.632
93.695 1.353
93.796 1.385
93.897 1.017
93.998 1.172
94.097 1.219
94.198 1.06
94.3 1.494
94.399 1.406
94.5 1.401
94.601 1.748
94.702 1.624
94.803 0.951
94.904 1.066
95.004 1.28
95.105 1.412
95.205 1.766
95.306 1.677
95.407 1.665
95.507 1.726
95.608 1.805
95.708 1.726
95.809 1.845
95.91 1.38
96.012 1.423
96.111 1.311
96.212 1.235
96.314 1.197
96.414 1.309
96.514 1.18
96.615 1.773
96.716 1.722
96.816 1.781
96.917 1.623
97.018 1.737
97.119 1.549
97.219 1.517
97.321 1.417
97.42 1.221
97.521 0.922
97.623 1.151
97.722 1.191
97.823 1.116
97.924 1.361
98.025 1.184
98.125 1.786
98.226 1.903
98.327 1.866
98.428 0.928
98.528 1.004
98.629 1.635
98.73 1.676
98.831 1.737
98.931 1.643
99.032 1.61
99.132 1.776
99.232 1.872
99.334 1.876
99.435 1.853
99.535 1.633
99.635 1.387
99.737 1.199
99.838 1.123
99.938 1.146
};
\addlegendentry{\small Harvester voltage}
\addplot [semithick, darkorange25512714]
table {%
0.00499999999999545 0.0136561392787276
0.106999999999999 0.06649643106749
0.207999999999998 0.0939557625921375
0.308999999999997 0.114350084936848
0.409000000000006 0.132211534425661
0.510000000000005 0.148186818767823
0.611000000000004 0.161573858290284
0.710000000000008 0.174865538937476
0.811999999999998 0.184780413943464
0.912999999999997 0.196138511979879
1.014 0.206874732770185
1.114 0.216618139212842
1.215 0.226632824577847
1.316 0.236237827613091
1.417 0.245621211762703
1.517 0.254554637333589
1.61799999999999 0.262962659682248
1.71900000000001 0.270543635583144
1.82000000000001 0.278247045583234
1.919 0.285600543836645
2.02 0.292958945172458
2.121 0.300082099513669
2.22199999999999 0.307005646412844
2.322 0.313740600519371
2.423 0.320390880947852
2.524 0.326773429125849
2.625 0.333052805233832
2.72499999999999 0.339549163668273
2.82600000000001 0.346066918864665
2.92700000000001 0.351862756316554
3.02800000000001 0.356833286967146
3.128 0.362597824784557
3.229 0.368617052934637
3.33 0.374181335680394
3.43000000000001 0.379942458513331
3.53100000000001 0.385490978985738
3.63200000000001 0.390736201838243
3.732 0.396458000942556
3.833 0.401170692328587
3.934 0.406420490993954
4.036 0.411951685860168
4.13500000000001 0.416568067908289
4.236 0.42161790422562
4.337 0.426419082445436
4.437 0.431492351221629
4.538 0.436757997448285
4.639 0.44147903726179
4.73999999999999 0.446339650182424
4.84100000000001 0.451330753227707
4.941 0.456037972587286
5.042 0.46092439698776
5.143 0.465961903232172
5.24299999999999 0.470264101699621
5.34400000000001 0.474695360415743
5.444 0.478939455164211
5.544 0.483494440570059
5.646 0.488424074425165
5.747 0.492657414755176
5.848 0.496930779059347
5.94800000000001 0.501230845858954
6.04900000000001 0.505234558797324
6.15000000000001 0.509105996229503
6.25 0.513138801292954
6.351 0.517249712540074
6.452 0.521689398446985
6.553 0.52548194584822
6.65300000000001 0.529660650983674
6.754 0.53409322495449
6.855 0.537811260380485
6.955 0.541668033447185
7.056 0.545728331796252
7.157 0.549048392008398
7.25700000000001 0.552787923714341
7.358 0.556251200978269
7.459 0.559946038232997
7.56 0.564073033181386
7.661 0.567390873167059
7.761 0.570997758363122
7.861 0.574875801548722
7.96300000000001 0.578351467525972
8.06400000000001 0.58206541600181
8.163 0.585641561041203
8.264 0.58931606389063
8.36499999999999 0.593000491175926
8.465 0.596591268094939
8.566 0.600208889442476
8.667 0.603682730755345
8.768 0.60689020732617
8.86799999999999 0.610293884534561
8.96900000000001 0.613873849383929
9.07000000000001 0.617460285362966
9.17100000000001 0.621272825135888
9.271 0.624766622480073
9.373 0.628205250685788
9.473 0.631566363669149
9.574 0.634820379185127
9.67400000000001 0.637922322608568
9.77500000000001 0.64099918772986
9.876 0.644377390974163
9.977 0.647480907086023
10.077 0.650773984204448
10.178 0.653546249485323
10.279 0.656804435919943
10.38 0.65963861540965
10.479 0.662639965739136
10.581 0.665625652266907
10.682 0.668876377635468
10.782 0.672051855352875
10.883 0.674730261212971
10.984 0.677442352229753
11.085 0.680581285559048
11.185 0.683390536533016
11.286 0.686392567953903
11.387 0.68921223339121
11.487 0.69218304019209
11.587 0.695260949684556
11.688 0.698273055735569
11.789 0.701020210629835
11.89 0.704024194673998
11.99 0.707151240411398
12.092 0.710057746560029
12.193 0.712900766770501
12.292 0.715771975800087
12.393 0.718739156411361
12.494 0.721773423174395
12.594 0.724478404045353
12.695 0.727281328025262
12.796 0.730413156205286
12.897 0.733117054432265
12.998 0.736130572034487
13.098 0.738603663732902
13.199 0.741360424245312
13.3 0.744041450607683
13.401 0.746958624111635
13.501 0.749598102936108
13.603 0.752484824313542
13.702 0.755072554326776
13.804 0.757552005380737
13.904 0.76026535603531
14.005 0.762775061605455
14.105 0.765318112678768
14.205 0.768048816005889
14.307 0.770872422362357
14.407 0.773245775241326
14.508 0.775746300146031
14.609 0.778092111173811
14.71 0.780830836222768
14.811 0.783263080724958
14.911 0.785643365695542
15.012 0.788540768795555
15.112 0.790719615424948
15.213 0.793267868709173
15.313 0.795871725133358
15.414 0.79821595711085
15.516 0.800595822543288
15.616 0.802908353529996
15.717 0.805313074143937
15.817 0.807894289478656
15.918 0.810209403924409
16.019 0.812656564008653
16.119 0.815076905265162
16.22 0.81748143506943
16.321 0.820063478867353
16.421 0.822573587476181
16.522 0.824775950899422
16.623 0.827271061566224
16.724 0.829557963649285
16.824 0.831779179531546
16.925 0.834316272027616
17.026 0.83683425291086
17.127 0.839027955482428
17.227 0.841297654128748
17.328 0.843408663061238
17.429 0.845722228898004
17.529 0.847689779291895
17.63 0.849893862660721
17.731 0.852198671698648
17.831 0.854620219876239
17.932 0.85671379397596
18.033 0.858924441533091
18.134 0.861000655241566
18.234 0.863176721411791
18.335 0.865239382847082
18.436 0.867197924186275
18.537 0.8695132918649
18.637 0.871436538440737
18.738 0.873532593158946
18.839 0.875744830543555
18.938 0.87756977926693
19.04 0.879805220058665
19.141 0.881692397770054
19.242 0.883592053878836
19.341 0.885726119272828
19.443 0.887479705523884
19.544 0.889720954813092
19.644 0.891902761904878
19.745 0.893771260122047
19.845 0.89592559688459
19.946 0.898268669223326
20.047 0.900280074123035
20.147 0.902445520605582
20.248 0.90435699346827
20.349 0.906230550406505
20.451 0.908401230052804
20.551 0.910195418953875
20.651 0.911981378552425
20.751 0.913894957547539
20.853 0.915807853104919
20.954 0.917822870605323
21.053 0.919815426989125
21.155 0.92151475502372
21.256 0.923669365631079
21.356 0.925591358407067
21.457 0.927208842385468
21.557 0.928869061613761
21.658 0.930927901788186
21.759 0.932741877304845
21.86 0.934452765868885
21.96 0.936206942495265
22.062 0.938226370774398
22.161 0.94016505408672
22.262 0.941753727216709
22.363 0.943452904964635
22.465 0.945332249185135
22.564 0.947024327791029
22.666 0.948783317930665
22.766 0.950831231486078
22.867 0.952751527064219
22.967 0.954581578397347
23.068 0.956244902847184
23.169 0.958067020682869
23.27 0.95969293597549
23.371 0.961468934442267
23.471 0.963140848505342
23.571 0.964862161916669
23.672 0.966627100084784
23.774 0.968375721353365
23.873 0.969904677683961
23.974 0.971599258712212
24.076 0.973686088697638
24.175 0.97529176323975
24.276 0.976934322346554
24.377 0.978544947609374
24.478 0.980063187415514
24.578 0.981588374878351
24.679 0.983135768449243
24.78 0.984569350435511
24.881 0.986380862578424
24.981 0.987784959398108
25.082 0.989373049635116
25.183 0.990788853346706
25.284 0.992390359664655
25.384 0.993996415187916
25.485 0.995642464206127
25.586 0.997133282067684
25.686 0.998671753877747
25.788 1.00026419834298
25.887 1.00178175214551
25.989 1.00330637858093
26.09 1.00454691527266
26.191 1.0062661828726
26.29 1.00775730671087
26.392 1.00921651047149
26.492 1.01066050539666
26.592 1.0122766824474
26.694 1.01383834661002
26.794 1.01546690404437
26.895 1.01689313481404
26.995 1.01829980771829
27.096 1.01977376808199
27.197 1.02110702606223
27.298 1.02240788687934
27.398 1.02398309765635
27.499 1.02526988803858
27.599 1.02678820381227
27.7 1.02801683597872
27.801 1.02923490069279
27.902 1.03055133425835
28.003 1.031658307136
28.103 1.03283564412919
28.204 1.03413292267236
28.304 1.03538913843031
28.405 1.03646271105753
28.506 1.03776233370996
28.607 1.0387958996907
28.708 1.04007266030185
28.809 1.04101593915659
28.909 1.04202142379881
29.009 1.043122445415
29.11 1.04414455921864
29.211 1.04526498192433
29.312 1.0466006663597
29.413 1.04743192879744
29.513 1.04848396147597
29.614 1.04940837624706
29.714 1.05054525830326
29.816 1.05154719971629
29.916 1.05263213589019
30.016 1.05368920591463
30.117 1.05473769584143
30.218 1.05581853752733
30.318 1.05693781799068
30.419 1.0580587485893
30.521 1.05898486675777
30.621 1.06005880704945
30.721 1.06127896427326
30.822 1.06220942537516
30.923 1.06323228425888
31.024 1.0641948960057
31.125 1.06505471569982
31.225 1.06623108696779
31.326 1.0672588587106
31.426 1.06816902175558
31.528 1.06902882838223
31.629 1.07005613939665
31.729 1.07085583244121
31.83 1.07171378434072
31.93 1.07253930521348
32.03 1.0736585432303
32.131 1.07430241707838
32.233 1.07506487795231
32.334 1.07582940278618
32.433 1.07660560044374
32.534 1.07758184717374
32.635 1.07859318766787
32.736 1.07950847262004
32.836 1.08018252133428
32.937 1.08122702079259
33.038 1.0818723336322
33.138 1.08274262691186
33.239 1.08342537655477
33.34 1.08427723905695
33.44 1.08495999314324
33.541 1.08565784463343
33.642 1.086476409809
33.743 1.08734306809104
33.843 1.08794848135535
33.944 1.08879836188976
34.045 1.08954460965356
34.145 1.09036272259692
34.246 1.09117615244081
34.347 1.09195641300227
34.448 1.09267211235707
34.548 1.09327638904783
34.649 1.09394490699677
34.75 1.0945839510881
34.85 1.0953061164564
34.951 1.09603879960008
35.052 1.09688028569608
35.153 1.0974487545953
35.254 1.09803391334544
35.354 1.0986966180032
35.455 1.09944191559189
35.556 1.10003492788458
35.656 1.10074944230864
35.757 1.10129487368682
35.857 1.10207974660188
35.959 1.10257495005921
36.059 1.10314349329529
36.159 1.10382636560205
36.26 1.10448100508617
36.362 1.10505851542823
36.462 1.10561808214585
36.562 1.10616887734793
36.663 1.10672870883201
36.764 1.10742102081005
36.864 1.10792942199678
36.965 1.10844677865513
37.065 1.10928066470084
37.166 1.10992140535351
37.266 1.11032412437689
37.367 1.1109566837315
37.469 1.11142738197645
37.569 1.11221450836345
37.67 1.11279092689801
37.77 1.11345813944492
37.872 1.11394394194949
37.972 1.1145488848302
38.073 1.11502306271243
38.173 1.1154516609161
38.274 1.11602756611985
38.375 1.11666689512085
38.475 1.11722370128848
38.576 1.11768487145672
38.677 1.11835445849525
38.778 1.11880485845383
38.879 1.11933213222493
38.978 1.11969516215212
39.079 1.1201684977848
39.181 1.12041747838788
39.28 1.12081482395639
39.381 1.1212260392886
39.483 1.12170643736185
39.583 1.12209974418641
39.683 1.12253598448084
39.785 1.12294114413482
39.885 1.12326513044754
39.987 1.12365442022974
40.087 1.12412262611463
40.188 1.12467288433901
40.288 1.12506004103571
40.388 1.1255131255665
40.49 1.12585891686395
40.591 1.12632500590773
40.691 1.12698326351442
40.791 1.12770731216347
40.893 1.12799490621069
40.994 1.12848690103782
41.093 1.12900654131055
41.194 1.12956831947626
41.296 1.13000364901854
41.395 1.1304392677262
41.496 1.13086627787655
41.597 1.13139080748356
41.698 1.13183513839717
41.799 1.13214760404825
41.899 1.13244245917391
42 1.1328237330451
42.101 1.13317404048394
42.201 1.13367246835201
42.302 1.13410055070827
42.403 1.13447762525698
42.504 1.13482627897493
42.604 1.13502003106258
42.705 1.13534244426564
42.806 1.1354617209071
42.905 1.13585443946656
43.007 1.13616813399615
43.107 1.1365145677802
43.209 1.13658902361938
43.308 1.13688914785265
43.41 1.13725967070198
43.511 1.13747475634335
43.611 1.1378818323489
43.712 1.13812385894393
43.813 1.13848215353575
43.914 1.13877110776892
44.014 1.13888505755017
44.114 1.13911396295777
44.216 1.13942506396159
44.317 1.13986023151218
44.416 1.14018811111417
44.517 1.14063837728005
44.618 1.1409975046125
44.719 1.14138488882447
44.82 1.14165060456771
44.92 1.14193544544602
45.022 1.14235397134577
45.122 1.14251903030509
45.222 1.14288156573228
45.323 1.14315994762777
45.424 1.14339720578251
45.525 1.14375961058493
45.625 1.14413038028806
45.726 1.14446500030202
45.827 1.1447983300635
45.928 1.14498706592989
46.028 1.14523368220924
46.129 1.14538979608187
46.23 1.14561962760815
46.331 1.14578408273114
46.431 1.14606637513666
46.532 1.14623148854686
46.633 1.14651413221841
46.733 1.14670851388655
46.834 1.14690829145974
46.935 1.14736607937305
47.036 1.14734524686631
47.136 1.14762196076814
47.236 1.14788170628675
47.338 1.14785255158814
47.439 1.14798163301682
47.539 1.14817368254035
47.639 1.14836608459062
47.74 1.14875572763697
47.842 1.14902216288596
47.941 1.14942865135165
48.043 1.14949355835026
48.144 1.14952217633931
48.243 1.14976003151086
48.345 1.15004324752516
48.446 1.15037274294091
48.546 1.1506685752615
48.647 1.15080054695849
48.747 1.15107196212753
48.848 1.15145317379863
48.95 1.15158337887331
49.049 1.1517084953921
49.15 1.15199805327813
49.251 1.15228831635526
49.351 1.15247464254398
49.452 1.15269922706349
49.553 1.15269197607864
49.654 1.15304847414604
49.754 1.15322575276187
49.855 1.15329518346964
49.957 1.15348131690103
50.057 1.15363194466815
50.157 1.15374917436843
50.258 1.153836224996
50.359 1.15393310783956
50.459 1.15430982063869
50.561 1.15436478471624
50.661 1.15437195130598
50.761 1.15450268721956
50.863 1.15474926047423
50.963 1.15484513876665
51.064 1.15489182820697
51.164 1.1550138192817
51.265 1.15513104139995
51.366 1.15531609837873
51.467 1.15536783492023
51.567 1.15553670005445
51.668 1.15575599660658
51.769 1.15593538186111
51.87 1.15600677786084
51.97 1.15615762327544
52.071 1.15644142856768
52.171 1.15667403748634
52.272 1.15677354469768
52.373 1.15684921553836
52.474 1.15688571451274
52.574 1.15720274343456
52.676 1.1574174294378
52.775 1.15755693680653
52.877 1.15767137694697
52.977 1.15781482225756
53.079 1.15797260221234
53.178 1.15810133253575
53.279 1.1582309297477
53.38 1.15840611402267
53.481 1.15861647051112
53.581 1.1587320263525
53.682 1.15875905333516
53.783 1.15895310403489
53.884 1.1591827678067
53.984 1.15929861281633
54.085 1.15953322774618
54.186 1.1596646726296
54.286 1.15989091407011
54.387 1.15987893277557
54.488 1.16006089590888
54.589 1.16001425415271
54.689 1.16016459999539
54.79 1.16017489180536
54.891 1.16033886623474
54.991 1.16040880572565
55.091 1.1605384933902
55.193 1.16058756796406
55.294 1.16078498842677
55.393 1.16091992010519
55.495 1.16105604372287
55.628 1.16178122389671
55.732 1.16173948902027
55.833 1.16182007295314
55.934 1.16206062297392
56.035 1.16217328948122
56.135 1.16216526812525
56.236 1.16234504151335
56.337 1.16248821051386
56.438 1.16272138217368
56.538 1.16292165222941
56.639 1.16302612033951
56.739 1.16312127530336
56.84 1.16329395433251
56.941 1.16326597726187
57.042 1.16323394472827
57.142 1.16350235334429
57.242 1.16362059298678
57.344 1.163623622
57.445 1.16368549849065
57.546 1.1639114401453
57.646 1.16399020310149
57.747 1.16407405410376
57.847 1.16409787735237
57.949 1.16421254322157
58.049 1.16413542873331
58.149 1.16430864002674
58.251 1.16427318013617
58.35 1.16429562645104
58.452 1.16422289671534
58.552 1.1644501745664
58.653 1.1645974537963
58.754 1.16457863171824
58.854 1.16456733302532
58.955 1.16483246407535
59.057 1.16484946911481
59.158 1.16489052713295
59.258 1.16502110301903
59.358 1.16527096571123
59.459 1.165528651245
59.56 1.16568219539356
59.659 1.16576386460434
59.761 1.16602492776387
59.862 1.16601164403994
59.962 1.16602250121268
60.063 1.1661629688422
60.164 1.16622128226882
60.264 1.16626746456878
60.365 1.16619916769593
60.466 1.16621657380052
60.567 1.16627538338612
60.667 1.16644246773567
60.768 1.16646371115292
60.868 1.16665733521159
60.969 1.16677856177399
61.071 1.16684991207751
61.171 1.16689432176074
61.272 1.16694215478887
61.372 1.1670045695712
61.472 1.16713404165599
61.573 1.1671402404957
61.675 1.16726084323635
61.775 1.16733679947185
61.876 1.16740360237769
61.976 1.16748148388984
62.077 1.16746391876672
62.179 1.16757008359198
62.278 1.16755316250978
62.379 1.16753804934395
62.48 1.16755720592476
62.581 1.16777460136831
62.682 1.1678830660231
62.782 1.1680052469552
62.882 1.16819723273589
62.983 1.16840807459059
63.084 1.1687062820091
63.185 1.16872008228011
63.286 1.16868556355456
63.386 1.1688641064379
63.487 1.16902860554745
63.588 1.16904422460723
63.689 1.16903654723815
63.789 1.16896877186975
63.89 1.1690498940807
63.991 1.16925879760384
64.091 1.16932621837394
64.192 1.16944968216743
64.293 1.16961177698399
64.393 1.16951117871543
64.494 1.1695809936395
64.595 1.16962978138742
64.695 1.16962660716064
64.796 1.16977718425914
64.896 1.16983724890885
64.997 1.16986877732429
65.098 1.16990410959537
65.199 1.17002698964618
65.3 1.17011972218469
65.401 1.17020447049407
65.501 1.17012662156552
65.601 1.17026615421453
65.702 1.17032944278652
65.803 1.17030953036947
65.904 1.17034908800186
66.004 1.17038548716106
66.105 1.17060924431869
66.206 1.17077079802959
66.306 1.17078948096298
66.408 1.17080139315732
66.508 1.17083361914078
66.61 1.17093563474306
66.709 1.17102347433058
66.81 1.1711330527294
66.911 1.17123190178864
67.011 1.17124067860587
67.113 1.17120880455052
67.213 1.1713525842804
67.314 1.171495558198
67.414 1.17147077502827
67.515 1.17154884072873
67.616 1.17162542204177
67.716 1.17164495376382
67.817 1.17167285820563
67.918 1.17167171645586
68.019 1.1716880399205
68.12 1.17179916439028
68.22 1.17170379726921
68.321 1.17177574514817
68.422 1.17197377253443
68.523 1.1721054846988
68.624 1.17205101730835
68.723 1.17213513389529
68.824 1.17209222987776
68.925 1.17230186268139
69.027 1.17216276973049
69.126 1.17217367589709
69.227 1.17218950512938
69.328 1.17229374224247
69.429 1.17241507955897
69.529 1.17247324848033
69.63 1.17248210296159
69.731 1.17244272114476
69.831 1.17256760892771
69.932 1.17256103462897
70.033 1.17259360918991
70.134 1.17266529525958
70.233 1.17272178591174
70.335 1.17281269435422
70.435 1.17280201018787
70.537 1.17289702797782
70.637 1.1728718321382
70.738 1.17290930000878
70.838 1.17302902795407
70.939 1.17300397208104
71.04 1.17308781560589
71.14 1.17316786321813
71.241 1.17318353084747
71.342 1.17335429566848
71.443 1.17344712445991
71.543 1.17345048034589
71.644 1.1735035041529
71.744 1.17350477257368
71.845 1.17353942604046
71.946 1.17357968324752
72.047 1.17357178278849
72.148 1.17348400037492
72.248 1.17348153638715
72.348 1.17342545781768
72.45 1.17351597604922
72.551 1.17360716707862
72.651 1.17362656247827
72.752 1.17375224018058
72.853 1.1737551496579
72.954 1.17383795647666
73.054 1.17386647288783
73.155 1.17395003042376
73.255 1.17395879505359
73.356 1.1740843679775
73.457 1.17411294358076
73.558 1.17414425945023
73.659 1.17410702095812
73.759 1.17416001828572
73.859 1.17429501849601
73.96 1.17436922535106
74.061 1.17430792417408
74.162 1.17431643365286
74.262 1.17435294964004
74.363 1.17445001178929
74.464 1.17455095951869
74.564 1.17437773515024
74.665 1.17443602578592
74.766 1.17455258606618
74.867 1.17470587089794
74.967 1.17459799666938
75.068 1.17465896815886
75.168 1.17478624517165
75.27 1.17495625429228
75.369 1.17483424882692
75.471 1.17500292830587
75.572 1.17496268012791
75.672 1.17499729487724
75.773 1.17495733327619
75.874 1.17496921429913
75.974 1.17494150290824
76.074 1.17511782643048
76.175 1.17515110867964
76.276 1.17520520295331
76.376 1.17524688250727
76.478 1.17519430690008
76.578 1.17522412827091
76.679 1.17523172013914
76.78 1.1751607592605
76.881 1.175164027643
76.981 1.17525629018831
77.082 1.17533515933065
77.183 1.17534923944146
77.284 1.17535805862297
77.383 1.17542668290749
77.485 1.17542310755551
77.585 1.17544212647762
77.686 1.17544433613117
77.787 1.17550253657965
77.887 1.17549729208241
77.988 1.17554959805465
78.09 1.17552503877507
78.19 1.17571133672739
78.291 1.17562311078788
78.391 1.17565810221722
78.493 1.17571246592254
78.592 1.17579099343062
78.694 1.17578060317668
78.794 1.17584622661303
78.895 1.17592665765423
78.995 1.17596788294456
79.096 1.1760808596426
79.197 1.1760877714285
79.297 1.17609922882347
79.398 1.17622958160082
79.499 1.17630683182412
79.6 1.17634182747501
79.7 1.17637844934424
79.802 1.17625400968737
79.901 1.17637644325928
80.002 1.17639596937501
80.103 1.17645711565972
80.204 1.1765530495349
80.304 1.17671356544399
80.405 1.1768682834726
80.507 1.1769128054271
80.607 1.17694060339497
80.708 1.17676919847081
80.808 1.17675619942729
80.909 1.17669380110962
81.009 1.17678423564749
81.11 1.17677430938126
81.211 1.17684462994438
81.312 1.176877347135
81.412 1.17686887037949
81.513 1.17688134844194
81.614 1.17702796981606
81.715 1.17703042542397
81.815 1.17715201853205
81.916 1.17717341458887
82.017 1.17709317009991
82.117 1.17713414670674
82.218 1.17718333787977
82.319 1.17722264329195
82.42 1.17725060052458
82.521 1.17721906882792
82.62 1.17737212770312
82.721 1.17738512167704
82.822 1.17720671882987
82.923 1.17728299443597
83.023 1.17735877364316
83.124 1.17719409990476
83.225 1.17715322955625
83.325 1.17720988752989
83.427 1.17716759776777
83.527 1.17729079348692
83.628 1.17720514336233
83.728 1.17726503491882
83.829 1.17726435100387
83.93 1.17724310493693
84.031 1.17721998693463
84.131 1.17713014535834
84.232 1.17705635774454
84.333 1.17709644480127
84.433 1.17712261873167
84.535 1.17727310820776
84.635 1.17726602523913
84.736 1.17734390492755
84.836 1.17747145847122
84.937 1.17760091747649
85.038 1.1776589933512
85.138 1.177751812888
85.239 1.17765266626917
85.34 1.17775092929138
85.441 1.17775034379813
85.541 1.17772034885446
85.642 1.17782170933921
85.743 1.17776321895487
85.844 1.17775556127291
85.944 1.17783284777043
86.045 1.1779058689155
86.145 1.17784953063592
86.247 1.17784889779435
86.347 1.17774543228641
86.447 1.17790040624941
86.549 1.17792631208563
86.65 1.17782601604302
86.75 1.17783199715291
86.851 1.17779524660525
86.951 1.17780177673651
87.052 1.17781798539497
87.153 1.1777946195899
87.254 1.17788053344568
87.353 1.17783928330986
87.455 1.17778531791697
87.556 1.17784960645462
87.656 1.1779372574666
87.756 1.17793751551816
87.858 1.17801094853007
87.958 1.17793039580605
88.059 1.17807361332098
88.16 1.17813198475289
88.261 1.17816020767583
88.362 1.17823822177241
88.462 1.17819026808735
88.562 1.17820001713989
88.664 1.17824278718665
88.764 1.17817721960055
88.864 1.17819858942639
88.966 1.17833753728659
89.067 1.17816082074888
89.166 1.17828435251693
89.267 1.17823853969491
89.369 1.17826226176599
89.47 1.17836647906736
89.57 1.17832688430041
89.671 1.17837061554768
89.771 1.17839612821004
89.872 1.17834804195302
89.973 1.17848644789829
90.073 1.17840443609243
90.174 1.17836472588246
90.275 1.17836944934015
90.377 1.17828693480412
90.476 1.1782298930128
90.577 1.17836757509214
90.678 1.17828712930654
90.778 1.17834171911427
90.88 1.17833233886867
90.98 1.17834202627552
91.081 1.17836932890201
91.181 1.17840642648294
91.283 1.17845649414021
91.383 1.17843757035442
91.483 1.17850534144985
91.585 1.17849221955282
91.685 1.17858975641588
91.786 1.17839836436735
91.886 1.17849285708262
91.988 1.1784971988663
92.087 1.17857901189269
92.189 1.17850217159518
92.29 1.17848031926987
92.391 1.17843458186095
92.49 1.17852056697723
92.591 1.17852372121446
92.692 1.17867360328657
92.793 1.17876339832249
92.894 1.17863847936748
92.994 1.17870369968449
93.095 1.17864304462794
93.196 1.17868144485049
93.296 1.17877300690233
93.397 1.17865618977981
93.497 1.17876620472837
93.599 1.17897914088693
93.699 1.17883377379858
93.8 1.17882362645904
93.901 1.17878656267515
94.002 1.17875411304554
94.102 1.17871009707086
94.202 1.17873293404611
94.304 1.17862510186919
94.404 1.17866863581224
94.504 1.17870397470917
94.605 1.17863375787533
94.706 1.17880436273814
94.807 1.17861672696588
94.908 1.17859260926631
95.008 1.17863444145091
95.109 1.17856844586669
95.209 1.17860010397471
95.311 1.17870964193802
95.411 1.17870928714787
95.511 1.17883893785424
95.612 1.1788773233057
95.713 1.17887520030435
95.813 1.17886705915502
95.914 1.178913059665
96.016 1.17882380010372
96.115 1.17887431808128
96.216 1.1788841587592
96.318 1.17896556054604
96.418 1.17888388643153
96.518 1.17888718912081
96.619 1.17892416551075
96.72 1.17888802550557
96.82 1.17888437305926
96.921 1.17902781451028
97.022 1.17902601856275
97.123 1.17903672840476
97.223 1.1790626984033
97.325 1.17907761865231
97.424 1.17916046299868
97.525 1.1790941271986
97.627 1.17901272393006
97.726 1.17908957520696
97.828 1.17911039661525
97.928 1.17911862516059
98.03 1.17915365790329
98.129 1.17921997968836
98.23 1.17920876889647
98.331 1.17926080468103
98.432 1.17910653196822
98.532 1.17906264647738
98.633 1.17909544429641
98.734 1.17913853743284
98.835 1.179221464446
98.935 1.17923533903219
99.036 1.17911362875286
99.137 1.17922715427011
99.236 1.17924953490114
99.338 1.17932599988721
99.439 1.17939503545969
99.539 1.17940098610052
99.639 1.17931241522052
99.741 1.17932986915974
99.841 1.17930668904476
99.942 1.17938831131733
};
\addlegendentry{\small Buffer capacitor voltage}
\addplot [semithick, red]
table {%
-4.995 1.74
104.939 1.74
};
\addlegendentry{\small Target voltage (MCU)}
\end{axis}

\end{tikzpicture}

%% file: tikz/cdf.tex
\begin{tikzpicture}

\definecolor{crimson2143940}{RGB}{214,39,40}
\definecolor{darkgray176}{RGB}{176,176,176}
\definecolor{darkorange25512714}{RGB}{255,127,14}
\definecolor{forestgreen4416044}{RGB}{44,160,44}
\definecolor{lightgray204}{RGB}{204,204,204}
\definecolor{steelblue31119180}{RGB}{31,119,180}

\begin{axis}[
width=\figurewidth,
height=0.8\figureheight,
scale only axis,
legend cell align={left},
legend style={
  fill opacity=0.8,
  draw opacity=1,
  text opacity=1,
  at={(0.97,0.03)},
  anchor=south east,
  draw=lightgray204
},
tick align=outside,
tick pos=left,
x grid style={darkgray176},
xlabel={Time [\si{\second}]},
xmajorgrids,
yminorgrids,
xmin=-0.725550000000001, xmax=93.0,
xtick style={color=black},
y grid style={darkgray176},
ylabel={CDF},
ymajorgrids,
ymin=-0.029, ymax=1.049,
ytick style={color=black},
ytick = {0,0.25,0.5,0.75,1.0},
line cap = round
]
\addplot [thick, darkorange25512714,line cap = round]
table {%
0 0
3.565 0.02
4.032 0.04
4.91 0.06
5.574 0.08
6.026 0.1
6.248 0.12
7.001 0.14
8.108 0.16
8.475 0.18
8.846 0.2
11.973 0.22
12.121 0.24
13.297 0.26
13.954 0.28
15.316 0.3
17.015 0.32
17.442 0.34
17.479 0.36
18.094 0.38
18.736 0.4
18.755 0.42
18.816 0.44
19.351 0.46
19.879 0.48
20.057 0.5
20.918 0.52
21.325 0.54
21.479 0.56
21.603 0.58
23.823 0.6
24.133 0.62
24.314 0.64
24.689 0.66
24.773 0.68
24.986 0.7
29.486 0.72
30.054 0.74
32.757 0.76
33.043 0.78
33.833 0.8
35.359 0.82
35.395 0.84
36.414 0.86
44.027 0.88
49.961 0.9
56.442 0.92
56.562 0.94
62.778 0.96
89.069 0.98
89.376 1
93 1
};
\addlegendentry{Realistic \gls{mcu}, single-tone}
\addplot [thick, darkorange25512714, densely dashed, line cap = round]
table {%
0 0
3.73 0
48.808 0.02
48.98 0.04
49.363 0.06
49.428 0.08
49.552 0.1
49.561 0.12
49.674 0.14
49.687 0.16
49.706 0.18
49.723 0.2
49.757 0.22
49.947 0.24
50.918 0.26
51.083 0.28
51.527 0.3
51.619 0.32
51.695 0.34
51.874 0.36
52.333 0.38
52.622 0.4
52.651 0.42
52.997 0.44
53.818 0.46
54.313 0.48
55.055 0.5
55.877 0.52
56.444 0.54
56.655 0.56
57.194 0.58
57.508 0.6
58.084 0.62
58.213 0.64
58.405 0.66
58.583 0.68
58.616 0.7
58.845 0.72
58.873 0.74
58.885 0.76
59.554 0.78
59.614 0.8
59.86 0.82
60.102 0.84
60.114 0.86
60.295 0.88
60.303 0.9
60.525 0.92
62.225 0.94
62.487 0.96
74.843 0.98
82.879 1
93 1
};
\addlegendentry{Realistic \gls{mcu}, multi-tone}
\addplot [thick, forestgreen4416044,line cap = round]
table {%
0 0
3.73 0.02
4.158 0.04
4.459 0.06
4.725 0.08
4.781 0.1
5.268 0.12
6.038 0.14
6.389 0.16
6.442 0.18
6.699 0.2
7.371 0.22
8.023 0.24
8.112 0.26
9.33 0.28
9.398 0.3
9.776 0.32
9.93 0.34
10.333 0.36
11.336 0.38
11.582 0.4
13.749 0.42
13.877 0.44
15.316 0.46
16.339 0.48
17.454 0.5
17.732 0.52
17.975 0.54
20.641 0.56
21.039 0.58
21.736 0.6
21.841 0.62
22.038 0.64
22.071 0.66
22.255 0.68
23.193 0.7
23.633 0.72
24.205 0.74
24.395 0.76
27.638 0.78
27.935 0.8
29.865 0.82
30.517 0.84
31.564 0.86
33.313 0.88
33.568 0.9
36.571 0.92
36.914 0.94
45.437 0.96
50.569 0.98
56.051 1
93 1
};
\addlegendentry{Ideal \gls{mcu}, single-tone}
\addplot [thick, forestgreen4416044, densely dashed,line cap = round]
table {%
0 0
3.73 0
24.158 0.02
24.195 0.04
24.255 0.06
24.263 0.08
24.271 0.1
24.3 0.12
24.3 0.14
24.304 0.16
24.316 0.18
24.332 0.2
24.384 0.22
24.397 0.24
24.433 0.26
24.517 0.280
24.654 0.3
24.714 0.32
24.839 0.34
24.847 0.36
24.965 0.38
25.041 0.4
25.214 0.42
25.254 0.44
25.42 0.46
25.432 0.48
25.5 0.5
25.61 0.52
25.645 0.54
25.674 0.56
25.678 0.58
25.721 0.6
25.863 0.62
25.956 0.64
26.096 0.66
26.129 0.68
26.507 0.7
26.523 0.72
26.532 0.74
26.6 0.76
26.632 0.78
26.68 0.8
26.789 0.82
26.829 0.84
26.89 0.86
26.99 0.88
26.998 0.9
27.015 0.92
27.059 0.94
27.079 0.96
27.111 0.98
27.465 1
93 1
};
\addlegendentry{Ideal \gls{mcu}, multi-tone}
\end{axis}

\end{tikzpicture}